\pdfoutput=1

\documentclass[hidelinks,12pt]{article}

\usepackage{graphicx,amsfonts,psfrag,layout,subcaption,array,longtable,lscape,booktabs,dcolumn,natbib,amsmath,amssymb,amssymb,amsthm,setspace,epigraph,chronology,color, colortbl,caption,wasysym,threeparttable,adjustbox,authblk,bbm,rotating}
\usepackage[]{graphicx}\usepackage[]{color}
\usepackage[page]{appendix}
\usepackage{hyperref, url} 
\usepackage[section]{placeins}
\usepackage[linewidth=1pt]{mdframed}
\usepackage[margin=1in]{geometry} 

\title{Gender gaps in frontier entrepreneurship? Evidence from 1901 Oklahoma land lottery winners}
\author[ ]{Jason Poulos\thanks{\emph{Address for correspondence:} 180 Longwood Avenue, Boston, MA 02115. \emph{E-mail:} poulos@hcp.med.harvard.edu. \emph{Acknowledgements:} I thank Drew Kemp, Christopher Fong, Lillian Geerts, Honami Kobayashi, Andrew Lam, and Alexa Ocampo for excellent research assistance. The paper benefited from constructive comments by Aditya Dasgupta and participants of ``The Development of the American West'' Symposium hosted by the USC Bedrosian Center. Data and code are available at \url{https://github.com/jvpoulos/ok-lottery}.}}
\affil[ ]{Department of Health Care Policy, Harvard Medical School}
\date{\today}
\usepackage{xr}
\usepackage[bottom]{footmisc}

\usepackage{footmisc}
\DefineFNsymbols{mySymbols}{{\ensuremath\dagger}{\ensuremath\ddagger}\S\P
	*{**}{\ensuremath{\dagger\dagger}}{\ensuremath{\ddagger\ddagger}}}
\setfnsymbol{mySymbols}

\usepackage{tabularx}
\newcolumntype{Y}{>{\raggedleft\arraybackslash}X}

\definecolor{Gray}{gray}{0.9}

\newcommand{\captionfonts}{\scriptsize}

\makeatletter  
\long\def\@makecaption#1#2{%
	\vskip\abovecaptionskip
	\sbox\@tempboxa{{\captionfonts #1: #2}}%
	\ifdim \wd\@tempboxa >\hsize
	{\captionfonts #1: #2\par}
	\else
	\hbox to\hsize{\hfil\box\@tempboxa\hfil}%
	\fi
	\vskip\belowcaptionskip}

\newtheorem*{assumption*}{\assumptionnumber}
\providecommand{\assumptionnumber}{}
\makeatletter
\newenvironment{assumption}[2]
{%
	\renewcommand{\assumptionnumber}{Assumption #1}%
	\begin{assumption*}%
		\protected@edef\@currentlabel{#1}%
	}
	{%
	\end{assumption*}
}
\makeatother

\theoremstyle{definition}
\newtheorem{definition}{Definition}
\theoremstyle{remark}

\newcommand{\indep}{\perp \!\!\! \perp}

\newcommand{\E}{\mathrm{E}}

\newcommand{\possessivecite}[1]{\citeauthor{#1}'s (\citeyear{#1})}

\begin{document} 
 
\begin{singlespacing}
\maketitle  
\thispagestyle{empty}
\end{singlespacing}


\begin{abstract} 
	\begin{singlespace} 
\noindent 
The paper investigates gender differences in entrepreneurship by exploiting a large-scale land lottery in Oklahoma at the turn of the 20$^{\text{th}}$ century. Lottery winners claimed land in the order in which their names were drawn, so the draw number is an approximate rank ordering of lottery wealth. This mechanism allows for the estimation of a dose-response function, which relates each draw number to the expected outcome under each draw. I estimate dose-response functions on a linked dataset of lottery winners and land patent records, and find the probability of purchasing land from the government to be decreasing as a function of lottery wealth, which is evidence for the presence of liquidity constraints. I find female winners were more effective in leveraging lottery wealth to purchase additional land, as evidenced by significantly higher median dose-responses compared to those of male winners. For a sample of winners linked to the 1910 Census, I find that male winners have higher median dose-responses compared to female winners in terms of farm or home ownership. These results suggest that liquidity constraints may have been more binding for female entrepreneurs in the market economy.\\

\noindent \emph{Keywords:} Causal inference; Entrepreneurship; Gender; Liquidity constraints; Natural experiments
\end{singlespace}
\end{abstract}	

\pagebreak
\setcounter{page}{1} 

\section{Introduction}

According to \citet{turner1956significance}, ``free land meant free opportunities" for settlers on the western frontier. Did settlers respond differently to these opportunities conditional on their gender? Motivated by idea that wealth shocks ought to boost
entrepreneurship in the presence of liquidity constraints, the paper exploits a large-scale land lottery in Oklahoma to investigate the relationship between lottery wealth and the later entrepreneurship of lottery winners. The paper provides a window into gender constraints on economic opportunity on the frontier by further exploring whether there are gender differences in the entrepreneurial activities of winners following the lottery. 

The 1901 Oklahoma land lottery opened for settlement over two million acres of land that was formerly occupied by the Kiowa, Comanche, Apache, and Wichita tribes. While previous land lotteries in Georgia were administrated by the state and had strict eligibility criteria \citep[e.g.,][]{poulos2019land}, the Oklahoma lottery was administered by the federal government and any adult citizen was eligible to participate. Even though women could not legally vote at the time, they were able to participate in the Oklahoma lottery. However, eligibility was restricted to heads of household, which practically excluded married women.\footnote{Several female lottery winners reportedly had to forfeit their claim to land when they married soon after winning the Oklahoma lottery \citep{bleakley2010shocking}.} Female heads of household represented only 3 to 7\% of the heads of household in Oklahoma at the time of the lottery \citep{maguire20191889}.\footnote{These shares are estimated using samples from the 1890 Oklahoma Territorial Census and 1880 Midwest Census, respectively.}  Lottery winners were required to homestead this parcel for three years to formalize their claim, unlike the Georgia land lotteries, which had no homesteading requirement. 

The wealth shock produced by the Oklahoma lottery was substantial: \citet{bohanon1998costs} estimate a lower-bound mean value estimate of \$500 for  each 160-acre lot at stake in the lottery, which is roughly equal to the average annual income in the state in 1900. This is an approximate mean value, and some of the lots were thought to be much more valuable: newspaper accounts at the time of the drawing indicate the most valuable lots in the district of Lawton to be worth between \$20,000 to \$40,000, more than 40 times the average annual income. Lottery winners claimed land in the order in which their names were drawn; thus, the draw number is an approximate rank ordering of the amount of lottery wealth in terms of the option value of the land \citep{bleakley2010shocking}. The rank-order drawing motivates the estimation of a dose-response function that relates each value of the dose (i.e., draw number) to the average outcome under each value of treatment. When estimating dose-response functions, observed confounding can be adjusted by modeling the outcome as a function of the treatment and a propensity score, which is the probability of receiving a treatment given observed covariates \citep{imbens2000role,hirano2004propensity}. 

The paper estimates the relationship between lottery wealth and the later entrepreneurship of lottery winners, as proxied by purchasing land or completing a homestead patent. I link lottery winners to land patent records to measure whether they subsequently purchased land titles though the Land Act of 1820, or filed a homestead patent through the Homestead Act (HSA) of 1862, in the decade following the lottery.\footnote{The HSA also had a head-of-household requirement, which practically excluded married women from homesteading.} In addition, I backward-link the winners to the 1900 Census to measure their pretreatment characteristics, and then forward-link to the 1910 Census to measure winners' home or farm ownership in 1910. I hypothesize that winners with relatively low draw numbers gain more wealth in terms of the option value of the land, and are thus better able to overcome liquidity constraints and purchase additional land, file a homestead patent, or subsequently own a home or farm. 

A central question of the paper is whether the estimated dose-response functions differ with respect to lottery winners' gender. Descriptive evidence reveals gender differences in the observed probabilities of purchasing land or completing a homestead patent, with female winners less likely to engage in these entrepreneurial activities; however, after removing bias due to observed confounding, I find female winners to have a significantly higher median dose-responses than male winners. In terms of home or farm ownership in 1910, I find male winners to have a higher median dose-response than female winners. These findings suggest that for entrepreneurial activities in the market economy, liquidity constraints may have been more binding for female winners. 

The political economy literature has focused on the historical origin of gender differences in labor participation \citep{alesina2011fertility,alesina2013origins,giuliano2015role}. This strand of literature contends that the type of agricultural technology adopted by societies in the preindustrial period had persistent impacts on contemporary female labor force participation and fertility. \citet{alesina2011fertility}, for instance, show societies that traditionally practiced capital-intensive plough cultivation which relied on mainly on male workers --- as opposed to labor-intensive shifting cultivation that relied on both male and female workers --- have lower female participation in the workplace and in entrepreneurial activities. Historical agricultural technology is also shown to have persistent effects on fertility, sex ratios, and cultural attitudes toward women \citep{alesina2018traditional, NBERw23635,giuliano2020gender}. Qualitative evidence points to women on the western frontier as being distinctly entrepreneurial and  independent due to labor shortages and the physical demands of the frontier economy \citep{harris1984sex,flexner1996century}.

Another relevant literature in economics demonstrates the importance of wealth for entrepreneurship and interprets this relationship as evidence of liquidity constraints, which restrict the ability of individuals to exchange existing wealth for other assets  \citep[e.g.,][]{holtz1994sticking,fairlie2012liquidity}. A common empirical strategy in this literature is to determine whether an individual's wealth affects the probability of becoming an entrepreneur; if it does, then it is evidence that liquidity constraints are present. Exogenous wealth shocks can therefore help identify the presence of liquidity constraints \citep{holtz1994entrepreneurial}. This literature frequently uses self-employment as a proxy for entrepreneurship and focuses on samples of individuals that have experienced windfall gains in terms of lottery winnings or inheritances. \citet{lindh1996self} and \citet{blanchflower1998makes}, for instance, show that the probability of being self-employed increases when people receive windfall gains in the form of lottery winnings and inheritances.

Much like recent empirical studies focused on the role of capital constraints in the historical U.S. South \citep{bleakley2013up,gonzalez2017start} and in developing countries \citep[e.g.,][]{banerjee2017credit}, the present study focuses on the economic condition of capital-constrained, small-scale entrepreneurs. \citet{gates1996jeffersonian} argue that public lands policies such as the HSA were ineffective in building wealth for small farmers because they often faced binding capital constraints. \citet{de2000mystery} draws parallels between contemporary developing countries and the historical U.S. west, highlighting the role of formal property rights in land for transforming informal assets into collateral for loans. The outcomes used in the present study--- the probability of completing a homestead or purchasing a land patents, and home or farm ownership --- serve as proxies for entrepreneurship because they capture the willingness of individuals to take on risk, which is an important dimension of the environment for small entrepreneurs in developing economies \citep{bleakley2013up}. 

Risk-taking has long been recognized to be one of the essential characteristics of entrepreneurship \citep{knight1921risk}.\footnote{This view recognizes that entrepreneurs often cannot get access to sufficient capital from capital markets, which is especially true in developing economies \citep{evans1989some,evans1989estimated}. A competing view of \citet{schumpeter1934theory} is that modern capital markets in developed economies allow entrepreneurs to find capitalists to bear risks. \citet{levine2017smart} studies entrepreneurship in the context of developed economies, and takes on the Schumpeterian view of the nature of entrepreneurship, arguing that self-employment is an imperfect proxy for entrepreneurship since the measure includes individuals engaged in non-disruptive economic activity.} The experimental economics literature finds strong evidence for gender differences in risk taking: in the studies surveyed by \citet{eckel2008men} and \citet{croson2009gender}, women are observed to be more risk averse than men. For example, women tend to invest less in risky assets compared to men when playing investment game experiments \citep{charness2012strong}. \citet{gong2012gender} find that men are less risk averse than women in both matrilineal and patrilineal societies.

The paper is organized as follows: the section immediately below reviews the historical background of the Oklahoma land lottery in the context of the broader political economy of federal land policies, and also reviews related empirical literature; Section \ref{data} describes the data and record linkage procedure used in the study, provides summary statistics on the linked data, and results of balance tests; Section \ref{estimation} describes the estimation procedure for causal dose-response relationships; Section \ref{sec:results} provides empirical results; Section \ref{sec:conclusion} concludes.

\section{Historical background and related literature}\label{background}

Over the course of the 19$^{\text{th}}$ century, the public domain expanded westward as a consequence of diplomatic treaties and federal land policies such as the HSA, which opened for settlement hundreds of millions of acres of western frontier land. Any adult household head --- including women, immigrants who had applied for citizenship, and freed slaves following the passage of the Fourteenth Amendment---  could apply for a homestead grant of 160 acres of land, provided that they live and make improvements on the land for five years and pay a \$10 filing fee. 

The implicit goal of the HSA was to promote rapid settlement on the western frontier \citep{allen1991homesteading}. The sparse population meant that state and local governments competed with each other to attract migrants in order to increase land values and tax revenue \citep{poulos2019state}. Frontier governments offered migrants broad access to cheap land and property rights, unrestricted voting rights, and a more generous provision of schooling and other public goods \citep{engerman2005evolution}. Consistent with this view, \citet{poulos2021rnn} estimate that the long-run impact of homestead policy on public school spending is equivalent to 2.5\% of the total per-capita public school expenditures in 1929.

\possessivecite{turner1956significance} frontier thesis posits that the frontier acted as a ``safety valve'' for relieving pressure from congested urban labor markets in eastern states by attracting foreign migrants and workers, many of whom were the lower end of the skill distribution. The view of the frontier as a safety valve has been explored by \citet{ferrie1997migration}, who finds evidence in a linked census sample of substantial migration to the frontier by unskilled workers and considerable gains in wealth for these migrant workers. \citet{bazzi2020frontier} also demonstrate that frontier settlers were disproportionately illiterate and foreign-born. The authors argue that the frontier also attracted individualists who were more equipped to thrive in the harsh frontier conditions: counties with longer historical frontier experience exhibit lower contemporary property tax rates, and their citizens are more opposed to taxation and redistribution.\footnote{\citet{homola2022gendered} combine the measure of historical frontier experience from \citeauthor{bazzi2020frontier} with World War II enlistment rates to show that women residing in counties with more frontier experience enlisted in the War at lower rates, while men enlisted at similar rates.}

There were substantial barriers to entry to homesteading, and homesteaders took on enormous risk in the five years required to file a homestead patent. One of the most significant obstacles to entry was the need for capital to build a successful farming operation: contemporary writers estimated that potential homesteaders required \$600 to \$1000 to start a farm \citep{deverell1988loosen}. Indeed, the high cost associated with starting and maintaining a farm casts doubt on the safety valve hypothesis \citep{danhof1941farm}. Homesteading was a risky venture: over the period of 1910 to 1919, out of 604,092 homestead entries in the U.S., totaling over 128 million acres, only 384,954 (63.7\%) resulted in successful patents \citep{shanks2005homestead}.\footnote{At least part of the discrepancy between homestead entries and filings, however, may be explained by the use of fraudulent filers to stake a claim in a homestead, with no intention of completing the patent, in order to allow timber and mining companies to extract resources from the land \citep{gates1936homestead}.}

\subsection{Related empirical literature}

The paper speaks to previous to empirical literature that leverages lotteried wealth to identify average causal effects of wealth on individuals' outcomes. \citet{poulos2019land}, for example, exploits the first two Georgia land lotteries in 1805 and 1807 to estimate the effect of winning land on subsequent officeholding and wealth. While the study finds no causal effect of wealth on the probability of candidacy or holding office, quantile regression estimates show that winning a prize in the 1805 lottery conferred lottery winners near the median of the wealth distribution a \$170 increase in wealth, in terms of the estimated value of slaves owned in 1820. \citet{bleakley2013up} use a sample of 1832 lottery participants linked to the full-count 1850 Census to investigate the effects of lottery prize values on the long-run wealth distribution of lottery winners, and estimate that winning a prize in the 1832 Georgia lottery significantly increases 1850 total census wealth (i.e., combined slave and real-estate wealth) by \$200. The authors find that these gains were concentrated in the middle of the 1850 wealth distribution, rather than in the lower tail of the distribution where there is limited access to capital. Using the same linked sample, \citet{bleakley2016} finds no evidence of a treatment effect on the wealth, literacy, or occupational standing of lottery winners or their descendants. Using a small sample of Oklahoma lottery winners registered in Lawton and linked to both the 1900 and 1910 censuses, \citet{bleakley2010shocking} finds that winners with lower draw numbers were more likely to remain in Oklahoma.

Several empirical studies survey lottery winners in order to estimate the effect of lottery wealth on subsequent redistributive attitudes \citep{doherty2006personal}; consumption behavior and labor market earnings \citep{imbens2001estimating}; and bankruptcy filings \citep{hankins2011ticket}. In these studies, the assignment of lottery prizes is random, yet there is substantial nonresponse when surveying lottery winners to measure their attitudes or earnings, which yields a sample where the prize amount is not independent of pretreatment characteristics. The present study avoids nonresponse bias because the outcome variables are extracted from observed economic behavior such as the purchasing of land patents and from census data. However, sample bias may arise as a result of the process of linking winner records to land patent or census records with imperfect information. 

\subsection{Oklahoma land openings and the 1901 lottery}\label{lottery-background}

\begin{figure}[htbp]
	\begin{center}
		\includegraphics[width=\textwidth]{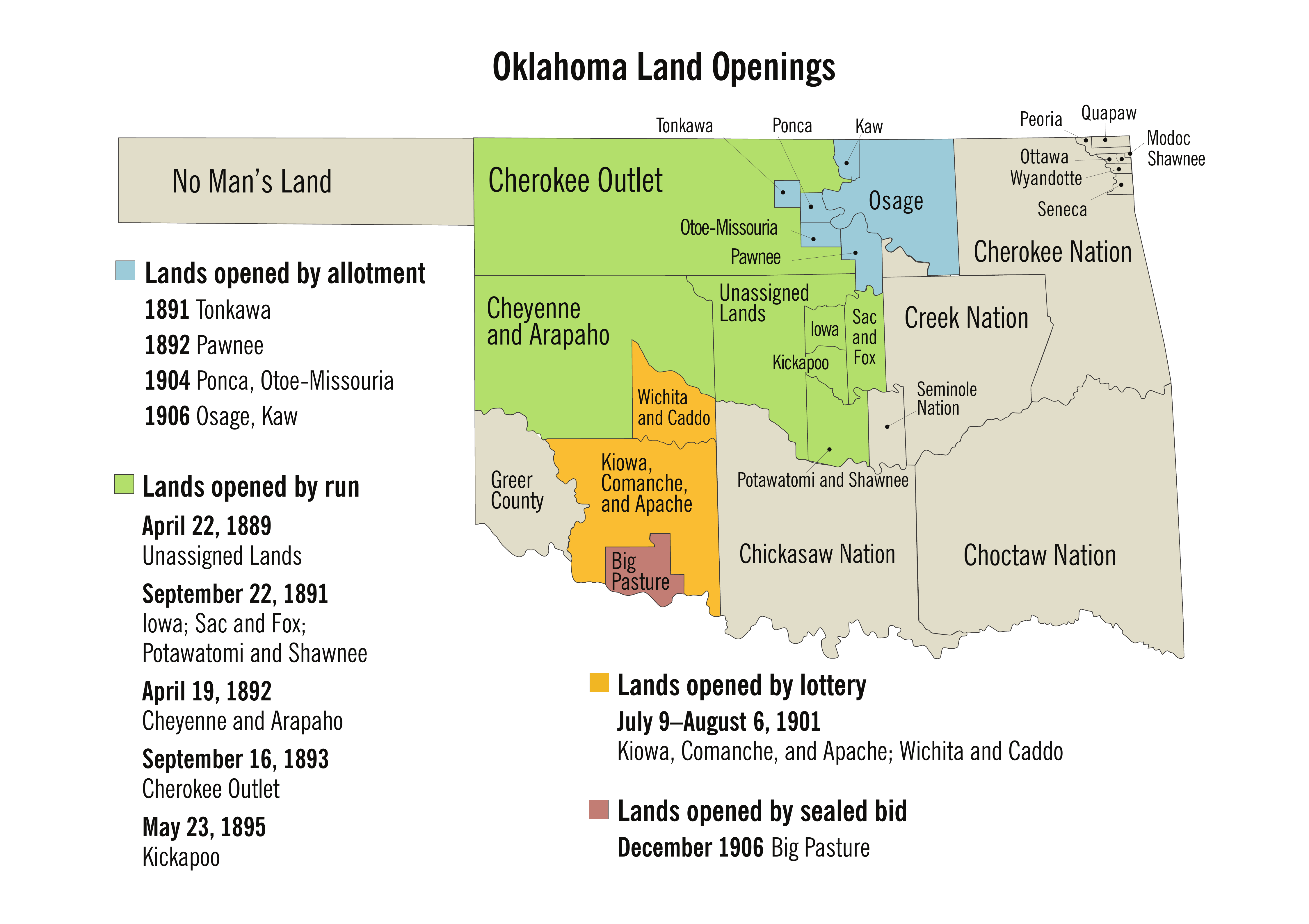} 
		\caption{Oklahoma land openings. Source: Oklahoma History Center (\url{http://www.okhistory.org/research/land}).}
		\label{fig:map}
	\end{center}
\end{figure}

In the years prior to Oklahoma's statehood in 1907, tribal lands in the central and western parts of Oklahoma were opened for settlement by the federal government to non-native American settlers. These lands were allocated by federal officials using four different mechanisms, as depicted in Figure \ref{fig:map}: allotment, land run, lottery, and sealed bid. In 1891, the Jerome Commission reached an agreement with several tribes that transferred ``surplus'' tribal land to the federal government that were to be opened to settlement under the HSA \citep{anderson1997}. Most of the tribal land in central Oklahoma, which were ceded to the federal government by the Creek and Seminole Indians following the Civil War, were opened by land run beginning in 1889. Land runs were literal races to claim surveyed lots, most the size of 160 acres, with fixed starting times and locations for participants \citep{bohanon1998costs}. Women who participated in the land runs had to be unmarried, widowed, or legally separated from their husbands \citep{maguire20191889}. As detailed below, the 1901 lottery randomly distributed claims to 160 acre land lots --- i.e., first drawers had the first right to claim land. Prairie land in southwestern Oklahoma known as the Big Pasture was opened by first-price sealed bid auction in 1906.  

About 3,250 square miles (2.08 million acres) of land in Kiowa-Comanche country --- formerly occupied by the Kiowa, Comanche, Apache, and Wichita tribes --- were instead opened for homestead settlement by lottery in July of 1901 and split into two districts: El Reno and Lawton. Lottery participants were required to be physically present at either the El Reno or Lawton land offices to register for the lottery, depending on whether they wished to claim land in El Reno or Lawton. El Reno had railroad access and an extra registration booth was opened to be utilized exclusively by women --- 8,000 women applied at this exclusive booth and an additional 2,000 women applied at the other booths in El Reno \citep{anderson1997}. Registration was open to heads of household at least 21 years old who did not already own more than 160 acres of land. Citizenship was not required for registration, even though the drawing was initiated by the federal government. Women were allowed to participate as long as they met the above requirements. 

About 170,000 people registered for the chance to win one of 13,000 lots of 160 acres in size. The lot size was not substantially smaller than 213 acres, which was the average acreage per farm in 1900 in the Oklahoma and Indian Territories at the time of the 1900 Agricultural census \citep{bohanon1998costs}.\footnote{Since Oklahoma lacked statehood in 1900, there is no county-level information on agricultural land values in the 1900 census. The 1910 Agricultural census data provides information on land values aggregated to the county level, while the land lots are tract-level, so it is not possible to map the association between land choice and land values. The historical archive of agricultural censuses is available at \url{https://agcensus.library.cornell.edu/}.}  Federal land officials randomly drew 6,500 envelopes containing participants' information (name, birthdate, and physical description) from each of two large hoppers --- one for participants who registered in Lawton and the other for those who registered in El Reno --- and envelopes were numbered as they were drawn \citep{anderson1997}. Results of the drawing were mailed to participants and posted in the newspapers. During the span of 60 days, lottery winners had the opportunity to ``stake [a] claim in turn" (i.e., in the order established by the drawing) on their desired land lot according to the number on the envelope \citep{watson1988}. 

The lands opened by lottery contained more fertile land than the arid land in western Oklahoma opened by run. \citet{bohanon1998costs} estimate the average value of the uncultivated lands opened by lottery was \$3.10 per acre, which leads to an estimate of \$496 for the 160-acre lot of uncultivated land --- roughly equal to the average annual income in 1900. Given this estimate and that the chance of winning a land lot was about 7.7\%, the expected value of participation was \$38.19. The average value of a cultivated Oklahoma farm of 160 acres is estimated to be between \$552 and \$2,786. Newspaper articles written at the time of the drawing estimate that the most valuable lots in Lawton were worth between \$20,000 to \$40,000.\footnote{One of these articles is ``Our Uncle Samuel's Lottery and Land Drawing in Oklahoma." \emph{Fort Morgan Times}, Volume XVII, Number 51, August 3, 1901. Available at \url{https://www.coloradohistoricnewspapers.org}.} These articles appeared days prior to the start of the land selection process for lottery winners, which is qualitative evidence that the location of the most valuable land lots were common knowledge. There was also weeks between the drawing and the registration of claims, so the winners would have a chance to go out into the territory and survey possible locations \citep{bleakley2010shocking}. 

\section{Data and record linkage} \label{data}

The analysis relies on lottery winner records, land patent records, and census data samples. First, I draw winner information from \possessivecite{anderson1997} list of Lawton winners and \possessivecite{watson1988} list of El Reno winners. The records contain information on winner name, place of residence, state or territory, draw number, and for the Lawton records, the land claimed by each winner. I combine the records of El Reno and Lawton winners, and drop winners with missing draws or draw numbers that occur more than two times, which yields a sample size of $n=13837$. In the sample, 7673 winners (55.5\%) registered in El Reno and had draw numbers 1 to 7731, where the Lawton records had 6164 winners (44.5\%) with draw numbers 1 to 6973. 

Table \ref{participant-gender} summarizes winner characteristics at the time of registration for the 1901 lottery. In the sample, 47.4\% of winners lived in Oklahoma Territory at the time of registration; 17.5\% lived in Kansas; and 10.5\% lived in Indian Territory. The three most frequent places lived at the time of registration were El Reno (3.8\%) and Oklahoma City (2.8\%), both in Oklahoma Territory, and Chickasha in Indian Territory (2.0\%). I code winners' gender by linking their first name to a public database of known female and male first names extracted from the modern census \citep{Maiaroto2014}. In total, 878 winners (6.3\%) are identified as having female first names. Compared to male winners, female winners are more likely to have registered in El Reno, and to have lived in El Reno and the Oklahoma Territory at the time of registration. An explanation for this discrepancy is that El Reno was easier to access by railroad and women were given an exclusive booth to register. 

\begin{table}
	\caption{Summary statistics for binary pretreatment covariates and post-treatment land patent outcomes, by gender.} \label{participant-gender}
	\begin{tabular}{lrrrrrr}
		\toprule
		\textbf{Variable} & $\mathbf{n_{\mathrm{male}}}$ & $\mathbf{\%_{\mathrm{male}}}$ & $\mathbf{n_{\mathrm{female}}}$ & $\mathbf{\%_{\mathrm{female}}}$ & $\mathbf{n_{\mathrm{all}}}$ & $\mathbf{\%_{\mathrm{all}}}$ \\ 
		\hline
		\emph{Gender} &  &  & &  &  &  \\
		Female & 0 & 0.0 & 878 & 100.0 & 878 & 6.3 \\ 
		\emph{District of registration} &  &  & &  &  &  \\ 
		Lawton & 5819 & 44.9 & 345 & 39.3 & 6164 & 44.5 \\ 
		\emph{State or territory} &  &  & &  &  &  \\ 
		\emph{(most frequent)}&  &  & &  &  &  \\ 
		Arkansas & 143 & 1.1 & 6 & 0.7 & 149 & 1.1 \\ 
		Illinois & 199 & 1.5 & 5 & 0.6 & 204 & 1.5 \\ 
		Indian Territory (I.T.) &  1372 & 10.6 & 86 & 9.8 & 1458 & 10.5 \\ 
		Iowa &  236 & 1.8 & 12 & 1.4 & 248 & 1.8 \\ 
		Kansas & 2297 & 17.7 & 127 & 14.5 & 2424 & 17.5 \\    
		Missouri &  1136 & 8.8 & 48 & 5.5 & 1184 & 8.6 \\   
		Nebraska &  137 & 1.1 & 6 & 0.7 & 143 & 1.0 \\ 
		Oklahoma Territory (O.T.) & 6024 & 46.5 & 538 & 61.3 & 6562 & 47.4 \\ 
		Texas & 1072 & 8.3 & 40 & 4.6 & 1112 & 8.0 \\ 
		\emph{Place of residence}&  &  & &  &  &  \\ 
		\emph{(most frequent)}&  &  & &  &  &  \\ 
		Anadarko, O.T. &162 & 1.2 & 8 & 0.9 & 170 & 1.2 \\ 
		Bridgeport, O.T. &104 & 0.8 & 8 & 0.9 & 112 & 0.8 \\ 
		Chickasha, I.T.&251 & 1.9 & 23 & 2.6 & 274 & 2.0 \\ 
		El Reno, O.T. &479 & 3.7 & 44 & 5.0 & 523 & 3.8 \\ 
		Enid, O.T. & 158 & 1.2 & 17 & 1.9 & 175 & 1.3 \\ 
		Fort Sill, O.T. &112 & 0.9 & 8 & 0.9 & 120 & 0.9 \\ 
		Guthrie, O.T. &157 & 1.2 & 15 & 1.7 & 172 & 1.2 \\ 
		Hobart, O.T. &186 & 1.4 & 10 & 1.1 & 196 & 1.4 \\ 
		Kansas City, Missouri &141 & 1.1 & 14 & 1.6 & 155 & 1.1 \\ 
		Kingfisher, O.T. &150 & 1.2 & 23 & 2.6 & 173 & 1.2 \\ 
		Mountain View, O.T. & 130 & 1.0 & 13 & 1.5 & 143 & 1.0 \\ 
		Oklahoma City, O.T. & 342 & 2.6 & 45 & 5.1 & 387 & 2.8 \\ 
		Perry, O.T. & 108 & 0.8 & 8 & 0.9 & 116 & 0.8 \\ 
		Shawnee, O.T. & 108 & 0.8 & 5 & 0.6 & 113 & 0.8 \\ 
		Weatherford, O.T. & 133 & 1.0 & 9 & 1.0 & 142 & 1.0 \\ 
		Wichita, Kansas & 114 & 0.9 & 14 & 1.6 & 128 & 0.9 \\ 
		\emph{Post-treatment outcome} &  &  & &  &  &  \\ 
		Land purchase & 1547 & 11.9 & 80 & 9.1 & 1627 & 11.8 \\ 
		Homestead & 980 & 7.6 & 28 & 3.2 & 1008 & 7.3 \\ 
		\bottomrule
	\end{tabular}
\end{table}

Second, I link winners to records of land patents under the Land Act of 1820, which allowed direct cash sales of public land from the General Land Office (GLO) to the public, and the HSA, issued within a decade of the 1901 drawing \citep{GLO}.\footnote{Patents issued under these two authorities account for three-quarters of the total land patents issued across all years.} The record linkage to land patent records is by exact match on full name (i.e., first, middle, and surname) and county/constituency (if available). Land patents are deeds transferring land ownership from the federal government to individuals, and provide information on the initial transfer of land titles from the federal government, including the patentee's name, date of patent, and location of land. 

The last two rows of Table \ref{participant-gender} shows that among all lottery winners, 11.8\% subsequently purchased land under the Land Act and 7.3\% filed a homestead patent under the HSA. There are substantial gender differences in the observed outcome rates: the male-female difference is 2.8\% for land purchase and 4.4\% for homestead patent. I calculate Pearson correlation coefficients and find positive correlations between draw number and the probability of land purchase, 0.05 ($p<0.001$), or homestead patent, 0.07 ($p<0.001$). The positive relationship between draw number and land purchase holds for the subgroup of female winners, 0.08 ($p=0.013$), but not for the homestead outcome, -0.01 ($p=0.827$). Unlike the causal estimates presented in Section \ref{sec:results}, these correlation coefficients include biases associated with differences in the covariates.  

In order to recover post-treatment outcomes from the 1910 Census, and pretreatment characteristics of the lottery winners from the 1900 Census, I backward-link the winners to the full-count 1900 Census, and then forward-link those matched to the full-count 1910 Census. I create a sample of heads of household from the full-count 1900 and 1910 censuses. The full-count census data have only have information that is useful for genealogical inquiry transcribed, such as name, county and state of residence, age, gender, race, and birthplace.\footnote{Access to the full-count data is granted by agreement between UC Berkeley, and the Minnesota Population Center (IPUMS USA). The Minnesota Population Center has collected digitized census data for 1790-1930 microdata collection with contributions from Ancestry.com and FamilySearch.} A team of undergraduate research assistants transcribed the remaining information for winners who were backward-linked to the 1900 Census and linked winners who were forward-linked to the 1900 Census. Using an automatic linking procedure described in Section \ref{linkage}, I successfully link 18.3\% of 1901 winners to the 1900 Census and 16.7\% of the linked sample to the 1910 Census. In comparison, \citet{bleakley2010shocking} manually link 24.3\% of Lawton winners to the 1900 Census and 33\% of their linked sample to the 1910 Census. 

Table \ref{participant-census} provides summary statistics on pretreatment covariates of lottery winners backward-linked to the 1900 Census ($n=2591$) and then forward-linked to the 1910 Census, for winners with non-missing farm and home ownership outcomes in 1910 ($n=397$). Since the census samples focus only on heads of household, only 1.7\% of lottery winners are female and a half of a percentage of winners linked to both the 1900 and 1910 censuses are female. Similar to the statistics in Table \ref{participant-gender}, the most frequent states of residence in 1900 for lottery winners linked to either census are Kansas, Missouri, and Texas.\footnote{The Indian Territory was not recognized as a state in either the 1900 or 1910 Census.} The 1900 Census provides a richer set of demographic characteristics on the linked lottery winners: the winners skewed young, with 34.3\% in the age group of 29 to 39 years old and 28.1\% in the age group of 18 to 28 years old. Ninety-two percent of the winners linked to the 1900 Census are white and 7.8\% are black. The 1910 Census provides information on two post-treatment outcomes of interest that capture entrepreneurial behavior: farm and home ownership. Among winners linked to both censuses, 50.6\% owned a farm and 66.5\% owned their own home in 1910. I calculate Pearson correlation coefficients and find no correlation between winners' draw number and the probability of owning a farm, 0.03 ($p=0.430$), or their own home, 0.03 ($p=0.481$).

\subsection{Balance tests}

I assess the plausibility of the key identifying assumption (Assumption \ref{a1} defined in Section \ref{estimation}) that treatment level and potential outcomes are independent conditional on the observed pretreatment covariates. While this assumption is not directly testable, we can test whether there are any imbalances in the treatment levels in terms of observed covariates. 

\begin{figure}[htb]
	\begin{center}
		\includegraphics[width=\textwidth]{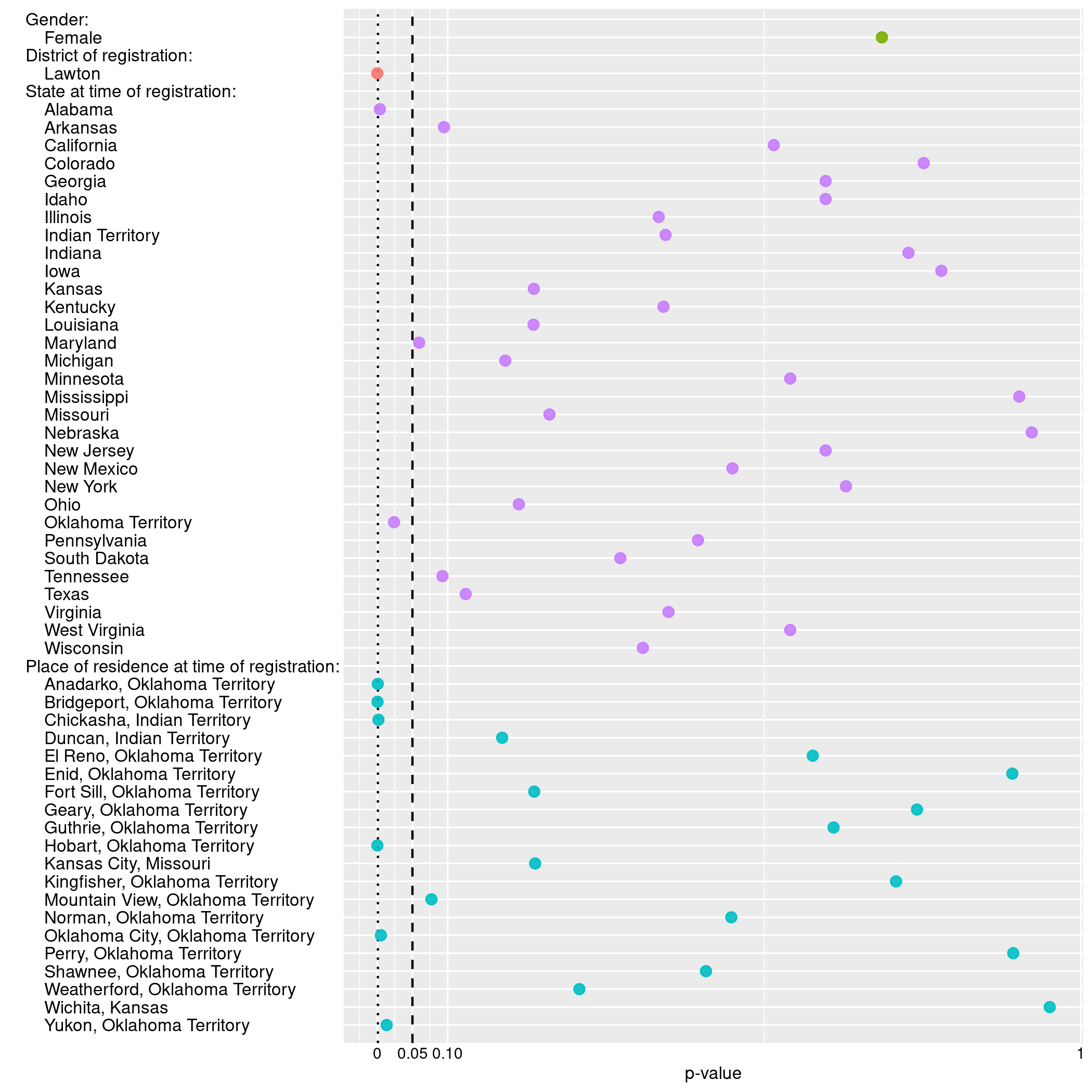}
		\caption{Balance in first draw number decile for El Reno and Lawton winners in terms of pretreatment covariates. $p$ values extracted from bivariate linear regressions of an indicator for draw number within first decile on each pretreatment covariate. The dotted vertical line is the Bonferroni corrected level of significance for 53 comparisons, $\alpha = 0.0009$, whereas the dashed vertical line is $\alpha = 0.05$. \label{balance-plot}} 
	\end{center}
\end{figure}

Figure \ref{balance-plot} reports balance in the first decile of draw number in terms of the covariates used to estimate the propensity score. Specifically, I create an indicator that takes the value of one if the draw number is within the first decile and consider the first decile as the reference treatment, since lower draw numbers correspond to more lottery wealth. Each point in the plot represents a $p$ value extracted from a bivariate linear regression of the first decile indicator on each pretreatment covariate. A statistically significant coefficient on the covariate implies an imbalance in first decile of draw number in terms of that covariate. 

While there is no evidence of imbalance in terms of winners' gender, there are imbalances in terms of winners' district of registration,  state, and of place of residence at the time of registration. The indicator for winners who registered at Lawton (rather than El Reno) is below the multiplicity-corrected level of significance (dotted line). Indicators for winners from Alabama and those living in several locations in the Oklahoma Territory (e.g., Anadarko, Bridgeport, Chickasha, Hobart, and Oklahoma City) are also significant at the multiplicity-corrected level. An indicator variable for winners living in Yukon in the Oklahoma Territory is significant at the conventional level of significance (dashed line). Most (but not all) of the Oklahoma Territory locations that exhibit imbalance are geographically closer to El Reno, so it is likely that winners residing in these locations would have registered at the El Reno land office. Examining the covariate distributions by draw number decile in Table \ref{participant-draw} reveals that the percentage of winners who registered in Lawton are evenly split for the first decile of draw number, but heavily imbalanced towards El Reno in the last decile. This is due to the fact that the El Reno records had more winners, and thus higher draw numbers. Winners living in the Oklahoma Territory locations that exhibit imbalance, whom presumably registered at the El Reno office, were proportionally more likely to draw numbers in the last two deciles. By including these covariates that exhibit imbalance in the propensity score model we are able to remove the resulting bias from the causal estimands. 

\section{Causal estimation} \label{estimation}

This study estimates the individual response (e.g., the probability of purchasing a land patent) that corresponds to specific values of continuous doses, i.e., random draw numbers. In the case of the Oklahoma lottery, a lower draw number corresponds to a higher dose in terms of the option value of the land. An analogy can be made to the studies of \citeauthor{doherty2006personal} and  \citeauthor{imbens2001estimating}, where a higher amount of prize corresponds to a higher dose.  

\subsection{Setup}

I follow the setup of \citet{hirano2004propensity} who propose estimating the entire dose-response function of a continuous treatment within the Neyman-Rubin potential outcomes framework \citep{neyman1923,rubin1990comment}. A sample of size $n$ is observed in which each lottery winner $i = \left\{1, ..., n \right\}$ has been assigned to treatment level $z_i \in \mathcal{Z}$, where $\mathcal{Z} = \{j = 1, 2, \ldots, J\}$ is the collection of possible treatment levels $j$ and $\mathbf{z}$ is a length-$n$ vector of assigned treatment levels. For each winner $i$, $y_i$ is the realized outcome and $\bf{x}_i$ is a vector of pretreatment covariates. Let the binary treatment indicator $d_{i}(j) = 1$ if $z_i = j$ and 0 otherwise. The potential outcome under treatment level $j$ is $y_i(j)$ and the realized outcome is $y_i = \sum_{j=1}^J d_i(j)y_i(j)$. The unit-level dose-response function is defined as the set of potential outcomes under each treatment level, $y_{i} (j): j \in \mathcal{Z}$. 

A common parameter of interest is the average dose response (ADR), or the average of the unit-level dose-response function, $\mu (j) = \E[y_{i} (j)]$. Another common parameter is the marginal treatment effect function \citep{michela2011nonparametric, kreif2015evaluation}, which captures the effect of increasing the treatment level by $\Delta j$ units on the expected potential outcome, $(\E[y_{i} (j)] - \E[y_{i} (j-\Delta j)])/\Delta j$. The focus in this study is on gender gaps, so interest lies in understanding the ADR function with respect to a subgroup, such as gender, defined by a subset of the covariate space, $\mathbb{C}$. A conditional marginal treatment effect function can similarly be estimated with respect to a subgroup. 
\noindent
\begin{definition}\label{cadr}
	{\em Conditional Average Dose Response (CADR). The average dose-response, treating $x \in \mathbb{C}$ as fixed.} %
	\begin{equation*}
	\mu_{j,x \in \mathbb{C}} = \E_{{\bf x} \in \mathbb{C}}[y_{i} (j)].
	\end{equation*} 
\end{definition}
\noindent
\begin{definition}\label{cmte}
	{\em Conditional Marginal Treatment Effect (CMTE). The marginal treatment effect function, treating $x \in \mathbb{C}$ as fixed.} %
	\begin{equation*}
	\left(\E_{{\bf x} \in \mathbb{C}}[y_{i} (j)] - \E_{{\bf x} \in \mathbb{C}}[y_{i} (j-\Delta j)]\right)/\Delta j.
	\end{equation*} 
\end{definition}

The causal parameters can be identified under unconfoundedness, or that adjusting for observed covariates ensures independence between treatment level and potential outcomes.
\begin{assumption}{1}{}\label{a1}
	Weak Unconfoundedness: $y (j) \indep z_i \mid {\bf x}$, \qquad  for all $j \in \mathcal{Z}$.
\end{assumption} 
\noindent 
The unconfoundedness assumption is weak because the conditional independence is assumed at each level of treatment rather than the joint independence of all the potential outcomes. This assumption implies that adjusting the conditional expectation of the outcome using information from the observed covariates is sufficient to remove bias from the estimate.

While it is not possible to directly test for unconfoundedness, the results of the balance tests in Figure \ref{balance-plot} evidence imbalance in terms of some of the covariates, in particular, winners' district of registration, state, and place of residence at the time of registration. Explicitly modeling the probability of receiving treatment as a function of covariates, and including the estimated propensity scores in the outcome model is expected to improve the consistency of the treatment effect estimates by eliminating any bias associated with differences in the observed covariates among groups of winners assigned to different treatment levels. Three additional estimation assumptions common in the causal inference literature are defined in Section \ref{assumptions} in the online supporting materials (SM). 

\subsection{Estimating dose-response curves}

The estimation of dose-response curves involves estimating both outcome and treatment models. This estimation approach is double-robust, which requires either the treatment model or outcome model to be correctly specified to ensure consistent estimation \citep{kang2007demystifying}. Misspecification of either the treatment or outcome model can lead to bias in the estimates of the dose-response function \citep[e.g.,][]{poulos2022targeted}. To avoid bias resulting from model misspecification,  I use the super learner \citep{van2007super,polley2010super,polley2011super} for estimating both outcome and treatment models in order to minimize the bias resulting from model misspecification. The super learner is an ensemble method that uses cross-validation to select the optimal weighted combination of many candidate classification or regression algorithms, where optimality is defined in terms of a pre-specified loss function \citep{kreif2015evaluation}.\footnote{In Tables \ref{ensemble-tab-treatment} to \ref{ensemble-tab-census}, I provide details of the candidate algorithms, training loss, and the weight assigned to each algorithm in the treatment and outcome model ensembles.}

The first step of estimating dose-response curves is to estimate the generalized propensity score $R = r(z, {\bf x})$, which is the conditional probability a winner is assigned to each treatment level, $\mbox{Pr}(z=j \mid {\bf x}_i)$. I estimate the generalized propensity score using the super learner estimate of winners' draw number given the pretreatment covariates included in Table \ref{participant-gender} for the analysis on land patent outcomes or Table \ref{participant-census} for the analysis on 1910 Census outcomes.  

The second step of estimating dose response curves is to estimate the conditional expectation of the outcome as a function of the treatment level and the propensity score, $\beta (z, R) = \E[y \mid z = j,  R = r]$ \citep{kreif2015evaluation}. I use super learner to model the outcome as a function of continuous treatment the estimated propensity scores. Then, the ADR function can be estimated by using the estimated parameters of the outcome model to predict the average potential outcome at each level of treatment. The CADR can be estimated by averaging over winners with covariates ${\bf x} \in \mathbb{C}$; that is, the average among those having ${\bf x} \in \mathbb{C}$. As \citet{hirano2004propensity} highlight, the regression function $\beta (z, R)$ does not itself have a causal interpretation, but comparing the ADR or CADR between different levels of treatment does have a causal interpretation.  

\section{Empirical results}\label{sec:results}

\begin{figure}[htb]
	\centering
	\begin{subfigure}{0.49\textwidth}
		\centering
		\includegraphics[width=\linewidth]{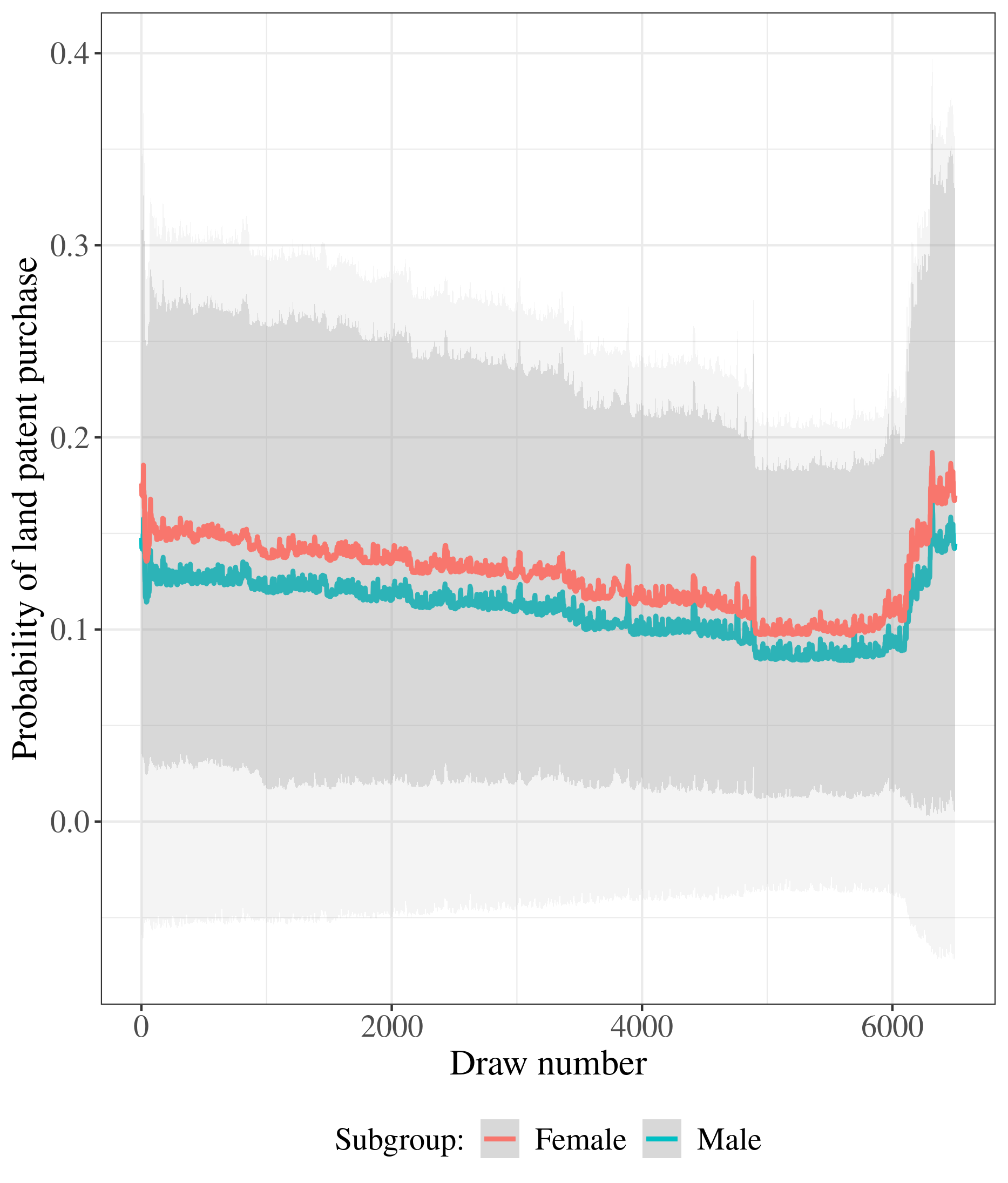}
		\caption{Average dose-response function}
	\end{subfigure}
	\hfill
	\begin{subfigure}{0.49\textwidth}
		\centering
		\includegraphics[width=\linewidth]{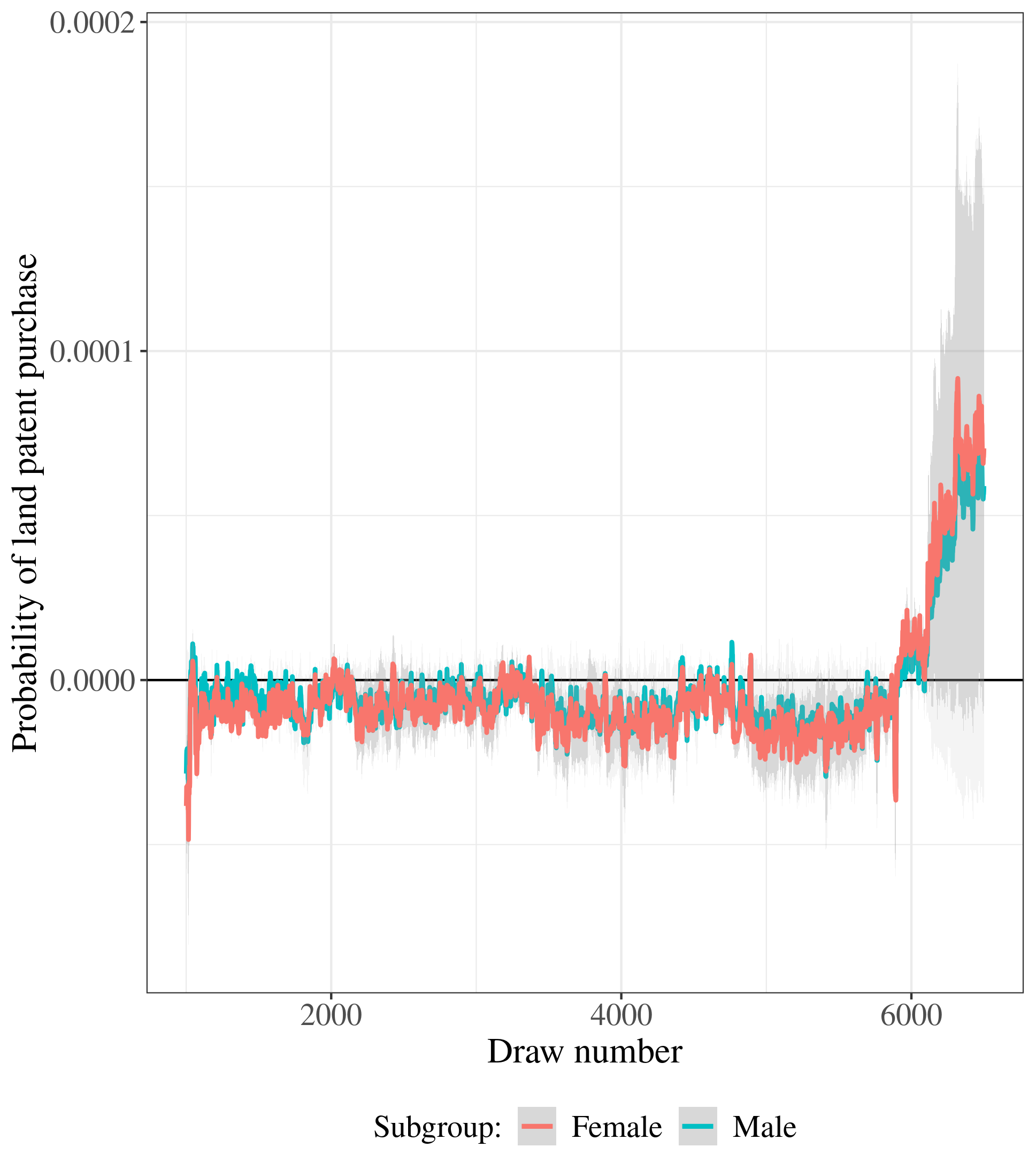}
		\caption{Marginal treatment effect function}
	\end{subfigure}
	\caption{Estimated CADR (a) and CMTE (b) for the land patent purchase outcome. The solid lines represent the estimated values by gender subgroup and the shaded regions represent 95\% confidence intervals calculated using the standard deviation of the predictions from all candidate algorithms in the super learner. The darker shaded regions correspond to the female subgroup.\label{sale-plots}} 
\end{figure}

I estimate the CADR (Definition \ref{cadr}) and CMTE (Definition \ref{cmte}) on the sample of El Reno and Lawton lottery winners for the land patents outcomes, by gender subgroup. Figure \ref{sale-plots}(a) plots the relationship between draw number (1 to 6500) and the probability of purchasing land for the subgroups of male and female winners. For both groups, the expected outcomes range from about 10\% to 20\%, depending on draw number, which is similar to the observed (i.e., unadjusted) land purchase rate for the sample, 11.8\%. The plot shows that for both male and female winners, the probability of land purchase is generally decreasing with draw number, although the relationship is non-monotonic since there is an uptick in the probability for winners with draw numbers over 6000. The negative relationship is consistent with the hypothesis that winners with relatively low draw numbers gain more wealth in terms of the option value of the land, and are thus better able to overcome liquidity constraints and purchase additional land. The non-monotonic relationship lacks a theoretical explanation since those with relatively high draw numbers got ``last dibs" on the available land. Since high-drawing winners were more likely to not claim lotteried land \citep{bleakley2010shocking}, an explanation might be that receiving last dibs motivated them to instead purchase land from the government.  

Visually comparing the CADR curves, there is a clear pattern of gender differences in terms of land patent purchases, with female winners actually more likely to purchase land compared to male winners. I conduct a Mann-Whitney U test to see if the mean ranks (which approximate the median) of the dose-response curves differ based on gender, resulting in a two-sided test $p$-value of less than 0.001, which is evidence against the null hypothesis that the CADR functions for male and female winners are equal. The Hodges-Lehmann estimate and 95\% confidence interval indicates that the median of the dose-response curve for female winners is 1.74 [1.68, 1.80] percentage points higher than the median for male winners.  

Figure \ref{sale-plots}(b) provides estimates of the CMTE, which captures the effect of increasing the draw number by $\Delta j=1000$ on the probability of land purchase for the gender subgroups. This parameter has a causal interpretation because it compares the expected potential outcomes under different levels of treatment; for instance, the difference in the expected potential outcomes under draw number 1001 and number 1, number 1002 and number 2, and so on. The CMTE point estimates are generally decreasing as a function of draw number until about the 6000$^{\text{th}}$ draw, which implies that increasing the draw number has a negative incremental effect on the expected probability of purchasing land. For instance, the incremental causal effect of increasing the draw number from 25 to 1025 decreases the probability of purchasing land by 0.002 [-0.00001, -0.004] percentage points for male winners and 0.003 [-0.001, -0.004] for female winners. While the negative causal effects are consistent with the liquidity constraint hypothesis, the finding of positive causal effects for winners on the right-tail of the curve lacks theoretical explanation. For instance, the incremental effect of increasing the draw number from 5058 to 6058 increases the probability of land purchase for both subgroups by 0.001 [0.001, 0.002] percentage points. These are very small incremental effects considering the observed land purchase rate is 11.8\%.

\begin{figure}[htb]
	\centering
	\begin{subfigure}{0.49\textwidth}
		\centering
		\includegraphics[width=\linewidth]{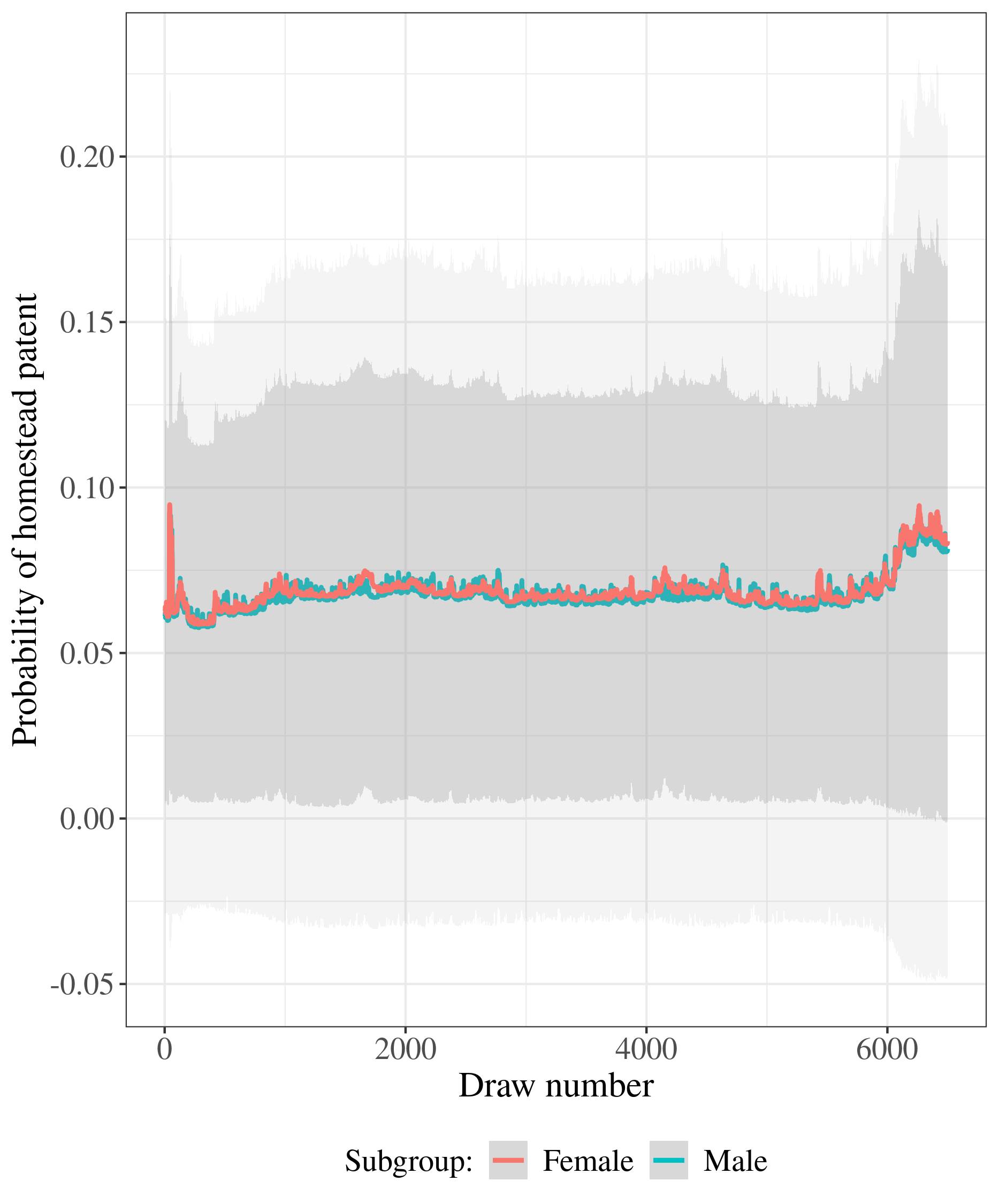}
		\caption{Average dose-response function}
	\end{subfigure}
	\hfill
	\begin{subfigure}{0.49\textwidth}
		\centering
		\includegraphics[width=\linewidth]{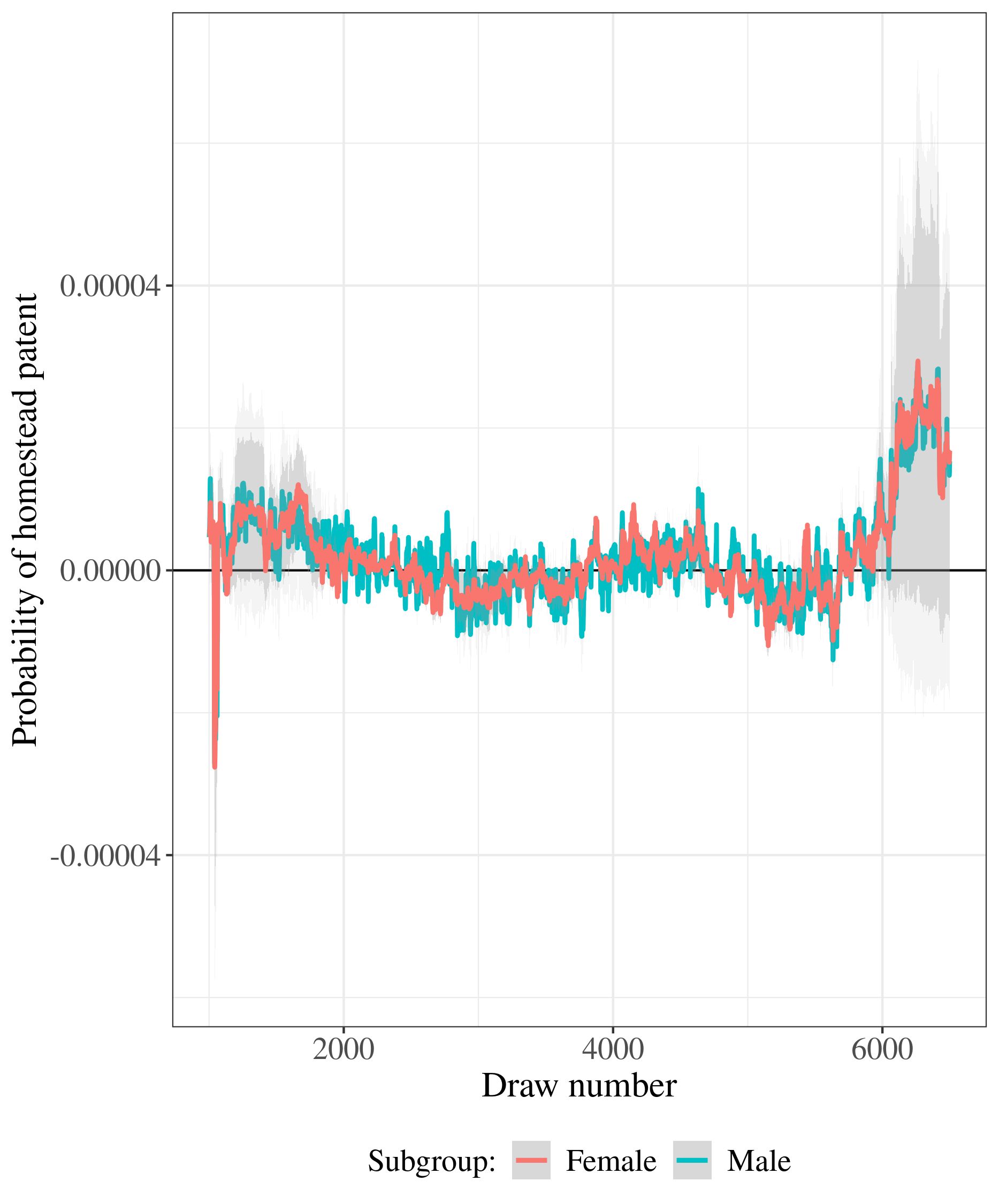}
		\caption{Marginal treatment effect function}
	\end{subfigure}
	\caption{Estimated CADR (a) and CMTE (b) for the homestead patent outcome. See notes to Figure \ref{sale-plots}.\label{homestead-plots}} 
\end{figure}

When the outcome is probability of completing a homestead patent within ten years following the 1901 lottery, the estimated CADR curves for male and female winners are relatively flat as a function of draw number, as shown in Figure \ref{homestead-plots}(a). For each group, the probability of completing a homestead ranges from about 5\% to 10\%, which is similar to the observed rate of 7.3\%. Results of a Mann-Whitney U test for difference between CADR curves demonstrate the median dose-response for female winners is 0.09 [0.08, 0.10] percentage points higher than the median for male winners --- a substantially smaller median difference compared to the land patent purchase outcome. Figure \ref{homestead-plots}(b), which plots the CMTE function, shows both negative or positive incremental effects, depending  on the draw number. For example, the incremental effect of moving from draw number 60 to 1060 yields for male winners a 0.002 [0.00002, 0.004] percentage point decrease in the probability of completing a homestead and a 0.001 [0.0003, 0.002] decrease for female winners. These are very small incremental effects compared to the observed rate completing a homestead for the sample. 

Figures \ref{farm-plots} and \ref{home-plots} report the estimated CADR functions on the probability of owning a farm or home in 1910, respectively, for the sample backward linked to the 1900 Census and forward-linked to the 1910 Census. Visually, there is no clear relationship between draw number and the expected probability of owning a farm or a home in 1910, for either gender subgroup. Male winners are more likely than female winners to own a farm in 1910 compared to female winners: the estimated median difference between the CADR functions is 1.54 [1.51, 1.57] percentage points higher for male winners. This is a relatively large difference, considering that the observed rate of farm ownership for the sample is 66.5\%. Male winners are also more likely to own a home in 1910 compared to female winners, with a median difference of 2.64 [2.58, 2.69], which is also a substantial difference considering the observed rate of home ownership in the entire sample is 50.6\%. The CMTE functions show incremental treatment effects on the probability of farm or home ownership that are practically zero. 

\section{Conclusion} \label{sec:conclusion}

There is substantial empirical evidence in economics pointing to (i.) gender differences in risk taking, which is an important element of entrepreneurship; and (ii.), the importance of wealth for entrepreneurship, which implies that liquidity constraints are present. However, not much is known about whether there are gender differences in the relationship between wealth and entrepreneurship, or how liquidity constraints impact entrepreneurial behavior conditional on gender.

The results of the paper can be interpreted as evidence of the presence of liquidity constraints among mostly white and male settlers on the western frontier. For both female and male lottery winners, the probability of purchasing land under the Land Act of 1820 subsequent to the lottery drawing is a declining function of the draw number. This finding indicates that those with more lottery wealth were more likely to buy land than those with less lottery wealth, and implies that liquidity constraints were present at the time of the lottery. Surprisingly, female winners have a higher dose-response curve than male winners, with a median difference of 1.74 [1.68, 1.80] percentage points, which suggests female winners were better able to leverage their lottery wealth into subsequent land purchases. This is a relative large median difference, considering the rate of land patent purchase is about 12\%. Female winners also have a higher dose-response in terms of completing a homestead, although the median difference is much smaller, 0.09 [0.08, 0.10]. For a much smaller sample of winners linked to the 1900 and 1910 censuses, I find that the median dose-response for male winners to be 1.54 [1.51, 1.57] percentage points higher than for female winners, and 2.64 [2.58, 2.69] higher in terms of the home ownership outcome. These are substantial differences relative to the observed rates of farm and ownership for the entire sample. 

Together, these findings suggest female winners were able to overcome liquidity constraints and either purchase additional land from the government or file a homestead patent. Both of these entrepreneurial activities were made possible by the federal government, with few restrictions on female participation. At the same time, there is evidence that men had the advantage in entrepreneurial activities in the market economy, such as home and farm ownership. These results suggest that liquidity constraints may have been comparatively more binding for female entrepreneurs in the market economy. 

\newpage
\begin{singlespace}
	\bibliographystyle{chicago}
	\bibliography{references}
\end{singlespace}	

\itemize
\end{document}


\begin{singlespacing}
\maketitle \thispagestyle{empty}
\tableofcontents \thispagestyle{empty}
\end{singlespacing}
	
\setcounter{figure}{0} \renewcommand{\thefigure}{SM-\arabic{figure}}
\setcounter{table}{0} \renewcommand{\thetable}{SM-\arabic{table}}
\setcounter{section}{0} \renewcommand{\thesection}{SM-\arabic{section}}
		
\pagenumbering{roman}
	
\pagebreak
\pagenumbering{arabic}

\section{Additional assumptions}\label{assumptions}

Two additional assumptions are needed to estimate the causal parameters. 
\begin{assumption}{2}{}\label{a2} 
	Stable unit treatment value assumption (SUTVA). Potential outcomes are well-defined and there is no interference between units; i.e., the potential outcome for one winner is a function of their treatment assignment and not the assignments of others.
\end{assumption} 
\noindent 
The first part of SUTVA eliminates spillover effects: the potential outcomes for a winner is caused only by their assigned treatment and not by treatments received by other winners. This no interference assumption is plausible for the treatment examined in this paper: a lottery winner's entrepreneurial behavior is likely unaffected by the draw number decile of another winner. The second part of SUTVA, which ensures that winners have the same number of potential outcomes, may be violated if there exists different versions of a treatment level. In the application, treatment level is well-defined by the draw number decile. The SUTVA implies consistency, or that the observed outcome corresponds to the potential outcome under the treatment level received:
\begin{assumption}{2}{}\label{a3}
	Consistency: $\text{If } z_i = j \text{ then } y_i (j) = y_i$. 
\end{assumption}
\noindent
Assumptions \ref{a2} and \ref{a3} are not directly testable. Estimating the propensity score assumes that there is a positive probability that a winner with a specific covariate pattern could be assigned to each $j$. 
\begin{assumption}{3}{}\label{a4}
	Positivity: $\mbox{Pr}(z = j \mid {\bf x}) > 0,~~~ \forall ~~ {\bf x} \in \mathbb{X} ~~\mbox{and}~~z \in \mathcal{Z}.$
\end{assumption}
\noindent
The positivity assumption restricts the sample to the covariate space were each winner has a positive probably to receive any treatment level. 

\section{Census record linkage} \label{linkage}

I follow the strategy of \citet{bleakley2010shocking} of backward-linking 1901 winner records to the 1900 Census sample and use the linked sample to link to the 1910 Census sample. This strategy allows for the use of pretreatment census variables, such as birthplace and birth year, which are used in the forward-linking classification model.

I follow closely the census sample linking procedure used by the Minnesota Population Center \citep{goeken2011,vick2011}. After preprocessing the character fields (e.g., removing non-alphabetic characters and standardization), I merge winner records to the 1900 Census samples by surname, first name initial, and state, and use surname, first initial, birthplace, and birth year when linking to the 1910 Census. To reduce the risk of false positives, I use census state files corresponding to the most frequent winner states of origin: Arkansas, Illinois, Iowa, Kansas, Missouri, Nebraska, Oklahoma, and Texas. Second, I form a training set by manually labeling records as a correct or incorrect match in a subset (i.e., 10\%) of the matched records. Third, I fit a a weighted ensemble of algorithmic classifiers on the training set using features that include Jaro-Winkler string distance in first name and city and binary variables indicating exact first name and city matches.\footnote{This approach differs from \citet{goeken2011}, who use a single support vector machine (SVM) classifier. The weighted ensemble method is expected to yield lower cross-validated classification error.}  Fourth, I use the ensemble fit on the unlabeled data, using a prediction threshold of 50\% to classify correct matches. Lastly, I exclude ambiguous multilinked records (e.g., multiple paired records for ``John Smith''). Details on the record classification ensemble weights and cross-validated training error is provided in Table \ref{ensemble-tab-link}.

\begin{table}[htbp]
	\begin{center}
		\caption{Error and weights for candidate algorithms for record classification ensemble.\label{ensemble-tab-link}} 
		\begin{tabular}{llcc}
			\hline
			\bf{Algorithm} & \bf{Parameters} & \bf{Training error} & \bf{Weight} \\ 
			\hline
			\rowcolor{Gray}
			\bf{1901 participants linked to 1900 Census} &  &  &  \\
			Generalized boosted regression (\texttt{gbm}) & default &0.066   & 0.789  \\
			Lasso regression (\texttt{glmnet}) & $\alpha=1$&  0.066 & 0.113\\ 
			GLM with elasticnet regularization (\texttt{glmnet}) &  $\alpha=0.25$ & 0.066 & 0.000 \\ 
			GLM with elasticnet regularization(\texttt{glmnet}) &  $\alpha=0.5$ &   0.066 &  0.000 \\ 
			GLM with elasticnet regularization (\texttt{glmnet})&  $\alpha=0.75$ & 0.066 &  0.000 \\ 
			Neural network (\texttt{nnet}) &  default & 0.142 & 0.024 \\ 
			Random forests (\texttt{randomForest}) & default & 0.104 &  0.000 \\ 
			Random forests (\texttt{randomForest})& $\mathrm{mtry}=1$ & 0.085 &  0.000 \\ 
			Random forests  (\texttt{randomForest})& $\mathrm{mtry}=5$  & 0.08 & 0.038 \\ 
			Random forests  (\texttt{randomForest}) & $\mathrm{mtry}=10$ & 0.025 &  0.000 \\ 
			Ridge regression  (\texttt{glmnet}) &  $\alpha=0$ & 0.086 & 0.033 \\ 
			\hline
			\rowcolor{Gray}
			\bf{Linked sample to 1910 Census} &  &  &  \\
			Generalized boosted regression (\texttt{gbm}) & default & 0.044  &  0.000  \\
			Lasso regression (\texttt{glmnet}) & $\alpha=1$&  0.043 &  0.988 \\ 
			GLM with elasticnet regularization (\texttt{glmnet}) &  $\alpha=0.25$ & 0.043 &  0.000 \\ 
			GLM with elasticnet regularization (\texttt{glmnet})&  $\alpha=0.5$ &   0.043 &  0.000 \\ 
			GLM with elasticnet regularization (\texttt{glmnet})&  $\alpha=0.75$ & 0.043 &  0.000 \\ 
			Neural network (\texttt{nnet}) &  default & 0.077 &  0.007 \\ 
			Random forests (\texttt{randomForest})& default & 0.056 &  0.000 \\ 
			Random forests (\texttt{randomForest})& $\mathrm{mtry}=1$ & 0.054 & 0.004 \\ 
			Random forests  (\texttt{randomForest})& $\mathrm{mtry}=5$  &0.053 &  0.000 \\ 
			Random forests  (\texttt{randomForest}) & $\mathrm{mtry}=10$ & 0.059 &  0.000 \\ 
			Ridge regression  (\texttt{glmnet})&  $\alpha=0$ & 0.051 &  0.000\\ 
			\hline
		\end{tabular} 
	\end{center}
	\footnotesize{Notes: cross-validated error and weights used for each algorithm in super learner ensemble. `Training error' is based on mean squared error for each algorithm during 10-fold cross-validation. `Weight' is the coefficient for the super learner, which is estimated using non-negative least squares based on the Lawson-Hanson algorithm. \textsf{R} package used for implementing each algorithm in parentheses. `GLM' stands for generalized linear model. $\mathrm{mtry}$ is the number
of predictors randomly sampled as candidates in each decision tree in random forests algorithm. $\alpha$
is a parameter that mixes L1 and L2 norms in the regularized regression.}
\end{table}

\section{Treatment and outcome model ensembles}

\begin{table}[htbp]
	\begin{center}
		\caption{Error and weights for candidate algorithms for treatment model ensembles.\label{ensemble-tab-treatment}} 
		\begin{tabular}{llcc}
			\hline
			\bf{Algorithm} & \bf{Parameters} & \bf{Training error} & \bf{Weight} \\ 
			\hline
			\rowcolor{Gray}
			\bf{Treatment model (for land patent outcomes)} &  &  &  \\
			Additive regression (\texttt{gam}) & $\mathrm{degree}=2$ & 4060723  & 0.000  \\
			Additive regression (\texttt{gam}) & $\mathrm{degree}=3$ & 4060723  & 0.000  \\
			Additive regression (\texttt{gam}) & $\mathrm{degree}=4$ & 4060723  & 0.000  \\
			Generalized boosted regression (\texttt{gbm}) & default & 4027647  & 0.568  \\
			Lasso regression (\texttt{glmnet}) & $\alpha=1$&  4052249 &  0.000 \\ 
			GLM (\texttt{glm}) &  default & 4060723 & 0.000 \\ 
			GLM with elasticnet regularization (\texttt{glmnet}) &  $\alpha=0.25$ & 4052277 & 0.000 \\ 
			GLM with elasticnet regularization (\texttt{glmnet})&  $\alpha=0.5$ &   4052318 & 0.000 \\ 
			GLM with elasticnet regularization (\texttt{glmnet})&  $\alpha=0.75$ & 4051893 & 0.000 \\ 
			Random forests (\texttt{randomForest})& default & 4033269 & 0.319 \\ 
			Random forests (\texttt{randomForest})& $\mathrm{mtry}=1$ & 4295872 & 0.000 \\ 
			Random forests  (\texttt{randomForest})& $\mathrm{mtry}=5$  &4054507 & 0.000 \\ 
			Random forests  (\texttt{randomForest}) & $\mathrm{mtry}=10$ & 4030585 & 0.000 \\ 
			Ridge regression  (\texttt{glmnet})&  $\alpha=0$ & 4059534 & 0.000\\ 
			Support vector machines (\texttt{svm})&  default & 4043854 & 0.111\\ 
			\rowcolor{Gray}
			\bf{Treatment model (for Census outcomes)} &  &  &  \\
			Additive regression (\texttt{gam}) & $\mathrm{degree}=2$ &  3632637  & 0.000  \\
			Additive regression (\texttt{gam}) & $\mathrm{degree}=3$ &  3632637  & 0.000  \\
			Additive regression (\texttt{gam}) & $\mathrm{degree}=4$ &  3632637 & 0.000  \\
			Generalized boosted regression (\texttt{gbm}) & default &  3476354  & 0.000  \\
			Lasso regression (\texttt{glmnet}) & $\alpha=1$&  3473983 &  0.000 \\ 
			GLM (\texttt{glm}) &  default & 3632637 & 0.000 \\ 
			GLM with elasticnet regularization (\texttt{glmnet}) &  $\alpha=0.25$ & 3472465 & 0.000 \\ 
			GLM with elasticnet regularization (\texttt{glmnet})&  $\alpha=0.5$ &   3471141 & 0.000 \\ 
			GLM with elasticnet regularization (\texttt{glmnet})&  $\alpha=0.75$ & 3474203& 0.000 \\ 
			Random forests (\texttt{randomForest})& default & 3609500 & 0.028 \\ 
			Random forests (\texttt{randomForest})& $\mathrm{mtry}=1$ & 3479833 & 0.000 \\ 
			Random forests  (\texttt{randomForest})& $\mathrm{mtry}=5$  &3643356  & 0.000 \\ 
			Random forests  (\texttt{randomForest}) & $\mathrm{mtry}=10$ & 3703836 & 0.000 \\ 
			Ridge regression  (\texttt{glmnet})&  $\alpha=0$ & 3465929 & 0.971 \\ 
			Support vector machines (\texttt{svm})&  default &  3631706 & 0.000\\ 
			\hline
		\end{tabular} 
	\end{center}
	\footnotesize{Notes: see notes to Table \ref{ensemble-tab-link}. $\mathrm{degree}$ is the smoothing term for smoothing splines for additive regression.}
\end{table}

\begin{table}[htbp]
	\begin{center}
		\caption{Error and weights for candidate algorithms for land patent outcome model ensembles.\label{ensemble-tab-land-patent}} 
		\begin{tabular}{llcc}
			\hline
			\bf{Algorithm} & \bf{Parameters} & \bf{Training error} & \bf{Weight} \\ 
			\hline
			\rowcolor{Gray}
			\bf{Outcome model (land purchase)} &  &  &  \\
			Generalized boosted regression (\texttt{gbm}) & default & 0.094   & 0.962  \\
			Lasso regression (\texttt{glmnet}) & $\alpha=1$&  0.100 & 0.000\\ 
			GLM with elasticnet regularization (\texttt{glmnet}) &  $\alpha=0.25$ & 0.100 & 0.000 \\ 
			GLM with elasticnet regularization(\texttt{glmnet}) &  $\alpha=0.5$ &   0.100 & 0.000 \\ 
			GLM with elasticnet regularization (\texttt{glmnet})&  $\alpha=0.75$ & 0.100 & 0.000 \\ 
			Random forests (\texttt{randomForest}) & default & 0.108 & 0.000 \\ 
			Random forests (\texttt{randomForest})& $\mathrm{mtry}=1$ & 0.108 & 0.000 \\ 
			Random forests  (\texttt{randomForest})& $\mathrm{mtry}=5$  & 0.124 & 0.000 \\ 
			Random forests  (\texttt{randomForest}) & $\mathrm{mtry}=10$ & 0.124 & 0.037 \\ 
			Ridge regression  (\texttt{glmnet}) &  $\alpha=0$ & 0.100 & 0.000 \\ 
			\rowcolor{Gray}
			\bf{Outcome model (homestead)} &  &  &  \\
			Generalized boosted regression (\texttt{gbm}) & default & 0.064   & 0.970  \\
			Lasso regression (\texttt{glmnet}) & $\alpha=1$&  0.066 & 0.000\\ 
			GLM with elasticnet regularization (\texttt{glmnet}) &  $\alpha=0.25$ & 0.066 & 0.000 \\ 
			GLM with elasticnet regularization(\texttt{glmnet}) &  $\alpha=0.5$ &   0.066 & 0.000 \\ 
			GLM with elasticnet regularization (\texttt{glmnet})&  $\alpha=0.75$ & 0.066 & 0.000 \\ 
			Random forests (\texttt{randomForest}) & default & 0.070 & 0.000 \\ 
			Random forests (\texttt{randomForest})& $\mathrm{mtry}=1$ & 0.070& 0.000 \\ 
			Random forests  (\texttt{randomForest})& $\mathrm{mtry}=5$  & 0.083 & 0.022 \\ 
			Random forests  (\texttt{randomForest}) & $\mathrm{mtry}=10$ &  0.083 & 0.007 \\ 
			Ridge regression  (\texttt{glmnet}) &  $\alpha=0$ & 0.066 & 0.000 \\ 
			\hline
		\end{tabular} 
	\end{center}
	\footnotesize{Notes: see notes to Table \ref{ensemble-tab-link}.}
\end{table}

\begin{table}[htbp]
	\begin{center}
		\caption{Error and weights for candidate algorithms for 1910 Census outcome model ensembles.\label{ensemble-tab-census}} 
		\begin{tabular}{llcc}
			\hline
			\bf{Algorithm} & \bf{Parameters} & \bf{Training error} & \bf{Weight} \\ 
			\hline
			\rowcolor{Gray}
			\bf{Outcome model (farm ownership)} &  &  &  \\
			Generalized boosted regression (\texttt{gbm}) & default & 0.253   & 0.000  \\
			Lasso regression (\texttt{glmnet}) & $\alpha=1$&  0.251 & 0.000\\ 
			GLM with elasticnet regularization (\texttt{glmnet}) &  $\alpha=0.25$ & 0.251 & 0.000 \\ 
			GLM with elasticnet regularization(\texttt{glmnet}) &  $\alpha=0.5$ &   0.251 & 0.000 \\ 
			GLM with elasticnet regularization (\texttt{glmnet})&  $\alpha=0.75$ & 0.251 & 0.000 \\ 
			Neural network (\texttt{nnet}) &  default & 0.250 & 0.925 \\ 
			Random forests (\texttt{randomForest}) & default & 0.298 &  0.074 \\ 
			Random forests (\texttt{randomForest})& $\mathrm{mtry}=1$ &0.298 & 0.000 \\ 
			Random forests  (\texttt{randomForest})& $\mathrm{mtry}=5$  &0.315 & 0.000 \\ 
			Random forests  (\texttt{randomForest}) & $\mathrm{mtry}=10$ &  0.316 & 0.000 \\ 
			\rowcolor{Gray}
			\bf{Outcome model (home ownership)} &  &  &  \\
			Generalized boosted regression (\texttt{gbm}) & default & 0.225   & 0.000  \\
			Neural network (\texttt{nnet}) &  default & 0.223 & 0.870 \\ 
			Random forests (\texttt{randomForest}) & default & 0.256 & 0.000 \\ 
			Random forests (\texttt{randomForest})& $\mathrm{mtry}=1$ & 0.254& 0.045 \\ 
			Random forests  (\texttt{randomForest})& $\mathrm{mtry}=5$  & 0.267 & 0.084 \\ 
			Random forests  (\texttt{randomForest}) & $\mathrm{mtry}=10$ &  0.267 & 0.000 \\ 
			\hline
		\end{tabular} 
	\end{center}
	\footnotesize{Notes: see notes to Table \ref{ensemble-tab-link}.}
\end{table}

\section{Summary and descriptive statistics}

\renewcommand{\tabcolsep}{4pt}
\begin{sidewaystable}[p]
	\caption{Summary statistics for binary pretreatment covariates and post-treatment outcomes\\ from El Reno and Lawton lottery winner records, $n=12759$, by draw number decile.} \label{participant-draw}
	\resizebox{\columnwidth}{!}{
		\begin{tabular}{l|rrrrrrrrrrrrrrrrrrrr|rr}
			\toprule
			\textbf{Variable}  & $\mathbf{n_{\mathrm{0-10}}}$ & $\mathbf{\%_{\mathrm{0-10}}}$ & $\mathbf{n_{\mathrm{10-20}}}$ & $\mathbf{\%_{\mathrm{10-20}}}$ & $\mathbf{n_{\mathrm{20-30}}}$ & $\mathbf{\%_{\mathrm{20-30}}}$ & $\mathbf{n_{\mathrm{30-40}}}$ & $\mathbf{\%_{\mathrm{30-40}}}$ & $\mathbf{n_{\mathrm{40-50}}}$ & $\mathbf{\%_{\mathrm{40-50}}}$ & $\mathbf{n_{\mathrm{50-60}}}$ & $\mathbf{\%_{\mathrm{50-60}}}$ & $\mathbf{n_{\mathrm{60-70}}}$ & $\mathbf{\%_{\mathrm{60-70}}}$ & $\mathbf{n_{\mathrm{70-80}}}$ & $\mathbf{\%_{\mathrm{70-80}}}$ & $\mathbf{n_{\mathrm{80-90}}}$ & $\mathbf{\%_{\mathrm{80-90}}}$ & $\mathbf{n_{\mathrm{90-100}}}$ & $\mathbf{\%_{\mathrm{90-100}}}$ & $\mathbf{n_{\mathrm{all}}}$ & $\mathbf{\%_{\mathrm{all}}}$  \\ 
			\hline
			\emph{Gender} &  &  & &  &  &  &  & &  &  &  & &  &  &  &  &  & & &  & & \\
			Female    & 91 & 6.6 & 79 & 5.7 & 73 & 5.3 & 89 & 6.4 & 96 & 6.9 & 96 & 6.9 & 83 & 6.0 & 100 & 7.2 & 89 & 6.4 & 82 & 5.9 & 878 & 6.3 \\ 
			\emph{District of} &  &  & &  &  &  &  & &  &  &  & &  &  &  &  &  & & & & & \\
			\emph{registration} &  &  & &  &  &  &  & &  &  &  & &  &  &  &  &  & & & & &   \\
			Lawton  & 688 & 49.7 & 662 & 47.9 & 646 & 46.7 & 687 & 49.6 & 688 & 49.7 & 675 & 48.8 & 687 & 49.6 & 685 & 49.6 & 672 & 48.5 & 74 & 5.3 & 6164 & 44.5 \\ 
			\emph{State or territory} &  &  & &  &  &  &  & &  &  &  & &  &  &  &  &  & & & & &   \\
			\emph{(most frequent)} &  &  & &  &  &  &  & &  &  &  & &  &  &  &  &  & & & & &  \\
			Arkansas  & 21 & 1.5 & 11 & 0.8 & 17 & 1.2 & 18 & 1.3 & 18 & 1.3 & 19 & 1.4 & 22 & 1.6 & 6 & 0.4 & 15 & 1.1 & 2 & 0.1 & 149 & 1.1 \\ 
			Illinois  & 24 & 1.7 & 21 & 1.5 & 23 & 1.7 & 22 & 1.6 & 19 & 1.4 & 20 & 1.4 & 22 & 1.6 & 28 & 2.0 & 18 & 1.3 & 7 & 0.5 & 204 & 1.5 \\ 	
			Indian Territory (I.T.)  & 137 & 9.9 & 145 & 10.5 & 134 & 9.7 & 139 & 10.0 & 167 & 12.1 & 141 & 10.2 & 152 & 11.0 & 160 & 11.6 & 166 & 12.0 & 117 & 8.5 & 1458 & 10.5 \\ 	
			Iowa  & 26 & 1.9 & 20 & 1.4 & 39 & 2.8 & 34 & 2.5 & 28 & 2.0 & 30 & 2.2 & 23 & 1.7 & 23 & 1.7 & 18 & 1.3 & 7 & 0.5 & 248 & 1.8 \\ 
			Kansas & 259 & 18.7 & 276 & 20.0 & 295 & 21.3 & 300 & 21.7 & 248 & 17.9 & 282 & 20.4 & 245 & 17.7 & 253 & 18.3 & 177 & 12.8 & 89 & 6.4 & 2424 & 17.5 \\ 
			Missouri  &  130 & 9.4 & 150 & 10.8 & 138 & 10.0 & 138 & 10.0 & 133 & 9.6 & 142 & 10.3 & 128 & 9.2 & 106 & 7.7 & 81 & 5.8 & 38 & 2.8 & 1184 & 8.6 \\ 	
			Nebraska  & 14 & 1.0 & 15 & 1.1 & 19 & 1.4 & 16 & 1.2 & 12 & 0.9 & 21 & 1.5 & 15 & 1.1 & 14 & 1.0 & 9 & 0.7 & 8 & 0.6 & 143 & 1.0 \\ 														
			Oklahoma Territory (O.T.) & 617 & 44.5 & 569 & 41.1 & 581 & 42.0 & 547 & 39.5 & 592 & 42.8 & 570 & 41.2 & 592 & 42.8 & 636 & 46.0 & 782 & 56.5 & 1076 & 77.8 & 6562 & 47.4 \\ 
			Texas  & 126 & 9.1 & 138 & 10.0 & 106 & 7.7 & 132 & 9.5 & 120 & 8.7 & 114 & 8.2 & 141 & 10.2 & 119 & 8.6 & 97 & 7.0 & 19 & 1.4 & 1112 & 8.0 \\ 
			\emph{Place of}&  &  & &  &  &  &  & &  &  &  & &  &  &  &  &  & & &  & &  \\
			\emph{residence}&  &  & &  &  &  &  & &  &  &  & &  &  &  &  &  & & & & &  \\
			\emph{(most frequent)}&  &  & &  &  &  &  & &  &  &  & &  &  &  &  &  & & &  & &  \\
		Anadarko, O.T. & 4 & 0.3 & 4 & 0.3 & 9 & 0.7 & 5 & 0.4 & 4 & 0.3 & 5 & 0.4 & 7 & 0.5 & 17 & 1.2 & 43 & 3.1 & 72 & 5.2 & 170 & 1.2 \\ 
Bridgeport, O.T. &0 & 0.0 & 1 & 0.1 & 4 & 0.3 & 3 & 0.2 & 1 & 0.1 & 1 & 0.1 & 2 & 0.1 & 6 & 0.4 & 23 & 1.7 & 71 & 5.1 & 112 & 0.8 \\ 
Chickasha, I.T.&12 & 0.9 & 15 & 1.1 & 20 & 1.4 & 16 & 1.2 & 31 & 2.2 & 24 & 1.7 & 28 & 2.0 & 27 & 1.9 & 52 & 3.8 & 49 & 3.5 & 274 & 2.0 \\ 
El Reno, O.T. &49 & 3.5 & 44 & 3.2 & 48 & 3.5 & 39 & 2.8 & 44 & 3.2 & 39 & 2.8 & 30 & 2.2 & 45 & 3.3 & 102 & 7.4 & 83 & 6.0 & 523 & 3.8 \\ 
Enid, O.T. & 18 & 1.3 & 25 & 1.8 & 17 & 1.2 & 21 & 1.5 & 17 & 1.2 & 17 & 1.2 & 18 & 1.3 & 16 & 1.2 & 18 & 1.3 & 8 & 0.6 & 175 & 1.3 \\ 
Fort Sill, O.T. &16 & 1.2 & 14 & 1.0 & 13 & 0.9 & 14 & 1.0 & 10 & 0.7 & 13 & 0.9 & 13 & 0.9 & 12 & 0.9 & 11 & 0.8 & 4 & 0.3 & 120 & 0.9 \\ 
Guthrie, O.T. &19 & 1.4 & 11 & 0.8 & 19 & 1.4 & 20 & 1.4 & 16 & 1.2 & 17 & 1.2 & 20 & 1.4 & 22 & 1.6 & 17 & 1.2 & 11 & 0.8 & 172 & 1.2 \\ 
Hobart, O.T. &4 & 0.3 & 1 & 0.1 & 2 & 0.1 & 0 & 0.0 & 2 & 0.1 & 4 & 0.3 & 3 & 0.2 & 1 & 0.1 & 45 & 3.2 & 134 & 9.7 & 196 & 1.4 \\ 
Kansas City, Missouri &11 & 0.8 & 19 & 1.4 & 23 & 1.7 & 15 & 1.1 & 18 & 1.3 & 18 & 1.3 & 18 & 1.3 & 17 & 1.2 & 9 & 0.7 & 7 & 0.5 & 155 & 1.1 \\ 
Kingfisher, O.T. &16 & 1.2 & 22 & 1.6 & 12 & 0.9 & 17 & 1.2 & 16 & 1.2 & 20 & 1.4 & 16 & 1.2 & 17 & 1.2 & 24 & 1.7 & 13 & 0.9 & 173 & 1.2 \\ 
Mountain View, O.T. &  8 & 0.6 & 6 & 0.4 & 9 & 0.7 & 9 & 0.7 & 6 & 0.4 & 3 & 0.2 & 9 & 0.7 & 15 & 1.1 & 28 & 2.0 & 50 & 3.6 & 143 & 1.0 \\ 
Oklahoma City, O.T. & 55 & 4.0 & 40 & 2.9 & 39 & 2.8 & 34 & 2.5 & 50 & 3.6 & 49 & 3.5 & 38 & 2.8 & 45 & 3.3 & 24 & 1.7 & 13 & 0.9 & 387 & 2.8 \\ 
Perry, O.T. &  12 & 0.9 & 11 & 0.8 & 8 & 0.6 & 20 & 1.4 & 15 & 1.1 & 10 & 0.7 & 16 & 1.2 & 11 & 0.8 & 8 & 0.6 & 5 & 0.4 & 116 & 0.8 \\ 
Shawnee, O.T. &  9 & 0.7 & 6 & 0.4 & 11 & 0.8 & 12 & 0.9 & 15 & 1.1 & 16 & 1.2 & 20 & 1.4 & 11 & 0.8 & 11 & 0.8 & 2 & 0.1 & 113 & 0.8 \\ 
Weatherford, O.T. & 18 & 1.3 & 11 & 0.8 & 10 & 0.7 & 12 & 0.9 & 13 & 0.9 & 14 & 1.0 & 9 & 0.7 & 14 & 1.0 & 19 & 1.4 & 22 & 1.6 & 142 & 1.0 \\ 
Wichita, Kansas & 13 & 0.9 & 19 & 1.4 & 17 & 1.2 & 16 & 1.2 & 16 & 1.2 & 14 & 1.0 & 9 & 0.7 & 9 & 0.7 & 12 & 0.9 & 3 & 0.2 & 128 & 0.9 \\ 
\emph{Post-treatment outcome} &  &  & &  &  &  \\ 
Land purchase &  163 & 11.8 & 159 & 11.5 & 155 & 11.2 & 138 & 10.0 & 150 & 10.8 & 122 & 8.8 & 137 & 9.9 & 94 & 6.8 & 183 & 13.2 & 326 & 23.6 & 1627 & 11.8 \\ 
Homestead & 75 & 5.4 & 80 & 5.8 & 94 & 6.8 & 84 & 6.1 & 76 & 5.5 & 84 & 6.1 & 93 & 6.7 & 81 & 5.9 & 129 & 9.3 & 212 & 15.3 & 1008 & 7.3 \\ 
			\bottomrule
		\end{tabular}
	}
\end{sidewaystable}
\renewcommand{\tabcolsep}{6pt}

\begin{table}[p]
	\caption{Summary statistics for binary pretreatment covariates\\ and post-treatment outcomes from winner records linked to 1900 and 1910 census data.} \label{participant-census}
	\begin{tabular}{lcccc} 
		\toprule
		&\multicolumn{2}{c}{Linked to 1900 Census} & \multicolumn{2}{c}{Linked to 1900 and} \\	
		& \multicolumn{2}{c}{($n=2591$)} & \multicolumn{2}{c}{1910 Census ($n=397$)}\\
		\textbf{Variable}  & $\mathbf{n}$ & $\mathbf{\%}$ & $\mathbf{n}$ & $\mathbf{\%}$ \\ 
		\hline
		\emph{Gender} &  & &  &   \\ 
		Female   &  44 & 1.7 & 2 & 0.5 \\ 
		\emph{State} &  & &  &   \\ 
		\emph{in 1900} &  & &  &   \\ 
		Kansas  &  1165 & 45.0  & 157 & 39.5 \\ 
		Missouri & 666 & 25.7   & 115 & 29.0 \\ 
		Texas  &  633 & 24.4 & 108 & 27.2 \\ 
		\emph{Race}  &  & &  &   \\ 
		Black  &  202 & 7.8 & 38 & 9.6 \\ 
		White  &  2386 & 92.1 &  358 & 90.2 \\
		\emph{Age} &  & &  &   \\ 
		18-28  &  728 & 28.1   & 98 & 24.7 \\   
		29-39  &  889 & 34.3  & 139 & 35.0 \\ 
		40-50  &  682 & 26.3 & 79 & 19.9 \\ 
		51-61 &  263 & 10.2    & 66 & 16.6 \\ 
		62-74  &  29 & 1.1  & 15 & 3.8 \\ 
		\emph{1910 Census}&  & &  &   \\ 
		\emph{outcome}&  & &  &   \\ 
		Own home   & -  & -&  264 & 66.5 \\   
		Own farm  & - & - &  201 & 50.6 \\  
		\bottomrule
	\end{tabular}
\end{table}

\section{Results for census outcomes}

\begin{figure}[htbp]
	\centering
	\begin{subfigure}{0.49\textwidth}
		\centering
		\includegraphics[width=\linewidth]{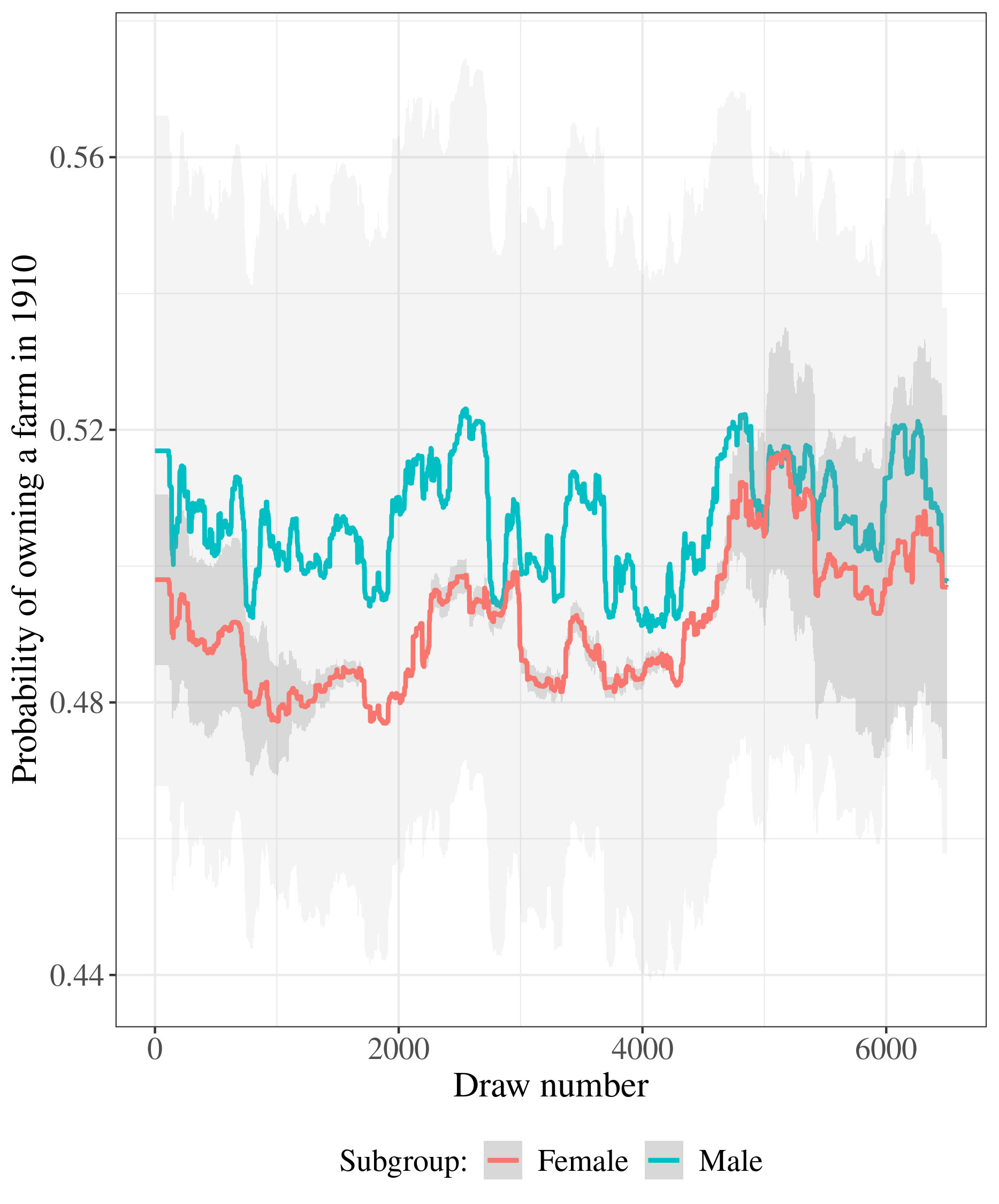}
		\caption{Average dose-response function}
	\end{subfigure}
	\hfill
	\begin{subfigure}{0.49\textwidth}
		\centering
		\includegraphics[width=\linewidth]{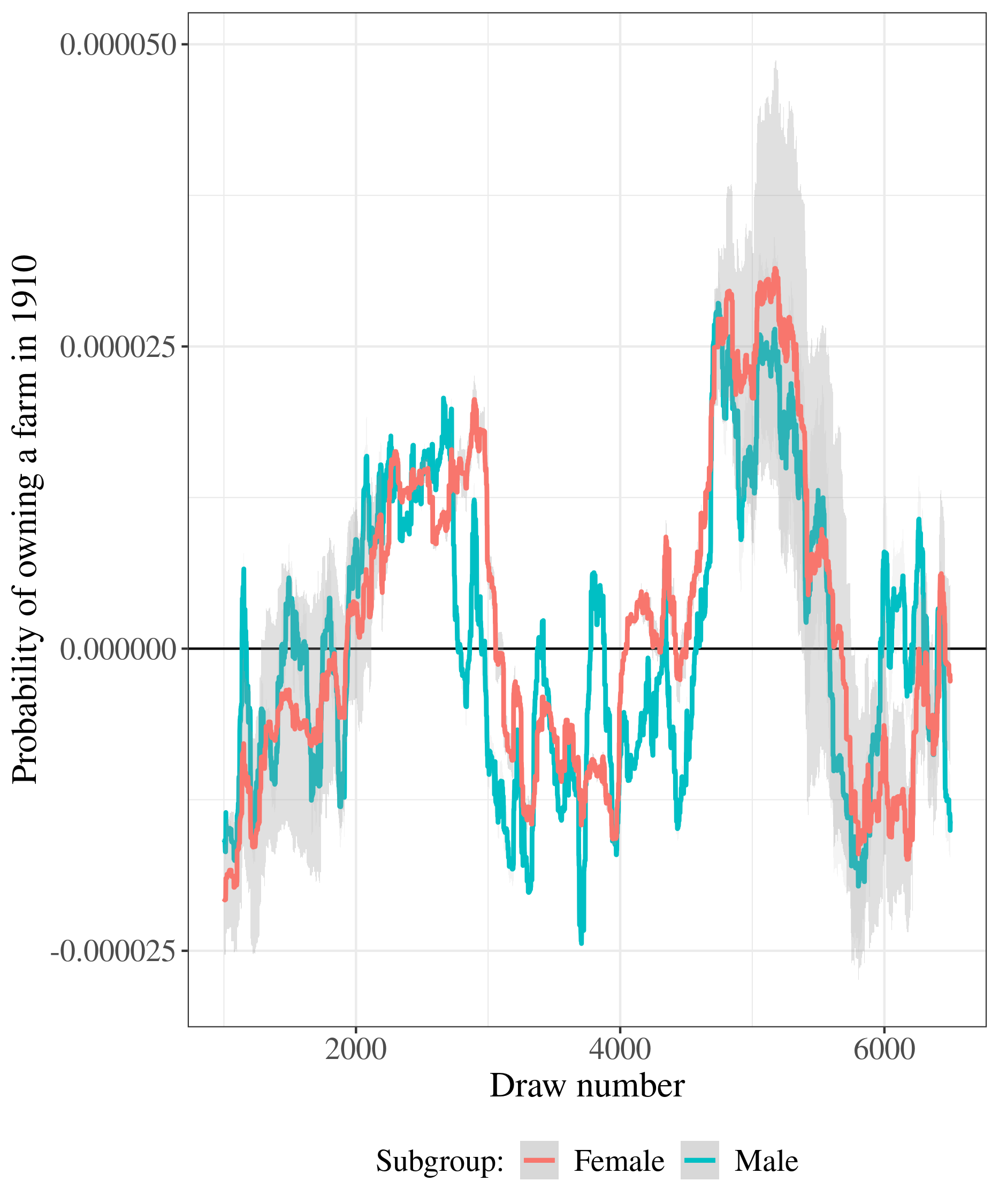}
		\caption{Marginal treatment effect function}
	\end{subfigure}
	\caption{Estimated average dose-response (a) and marginal treatment effect (b) on the probability of owning a farm. See notes to Figure \ref{sale-plots}.\label{farm-plots}} 
\end{figure}

\begin{figure}[htbp]
	\centering
	\begin{subfigure}{0.49\textwidth}
		\centering
		\includegraphics[width=\linewidth]{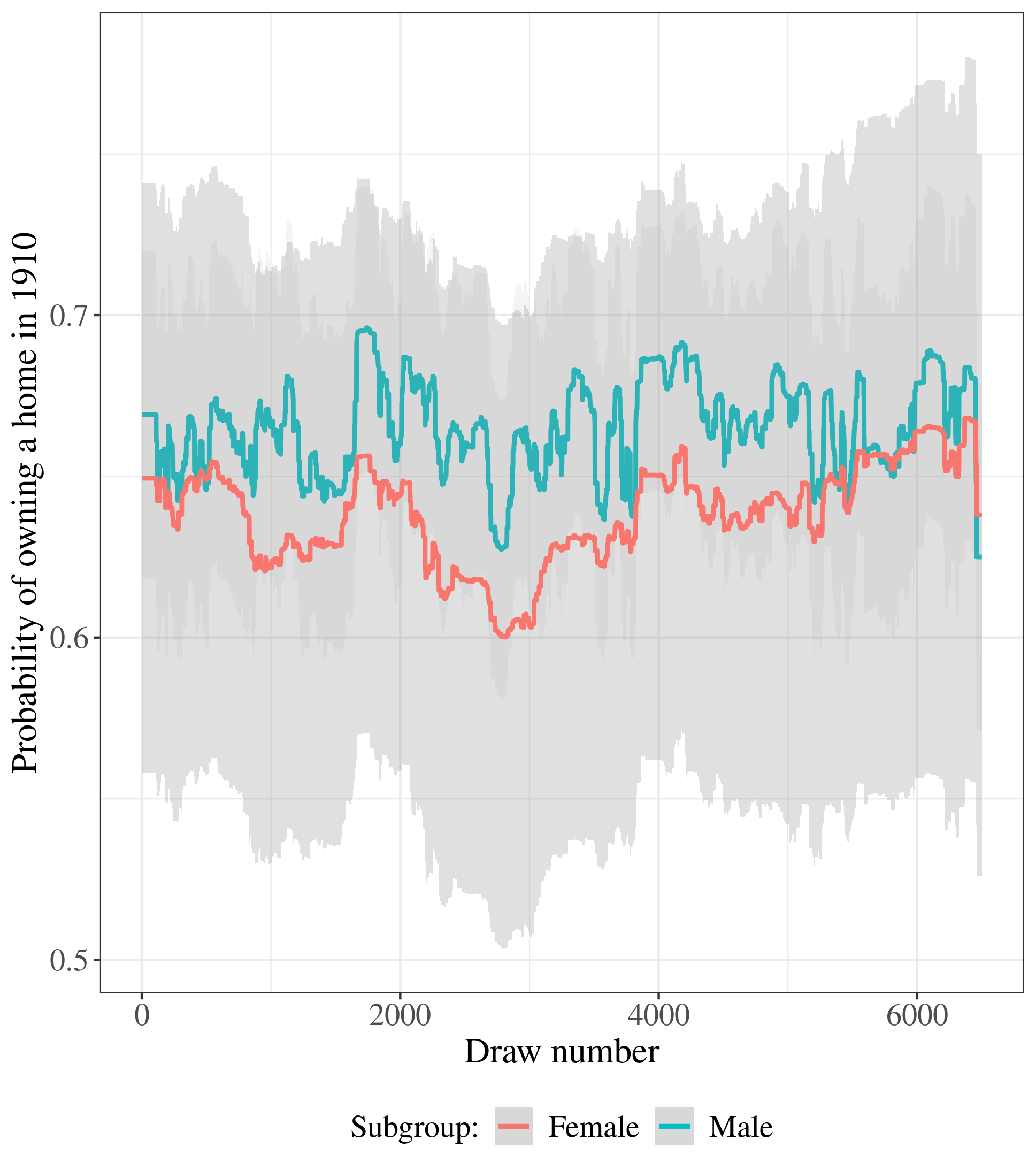}
		\caption{Average dose-response function}
	\end{subfigure}
	\hfill
	\begin{subfigure}{0.49\textwidth}
		\centering
		\includegraphics[width=\linewidth]{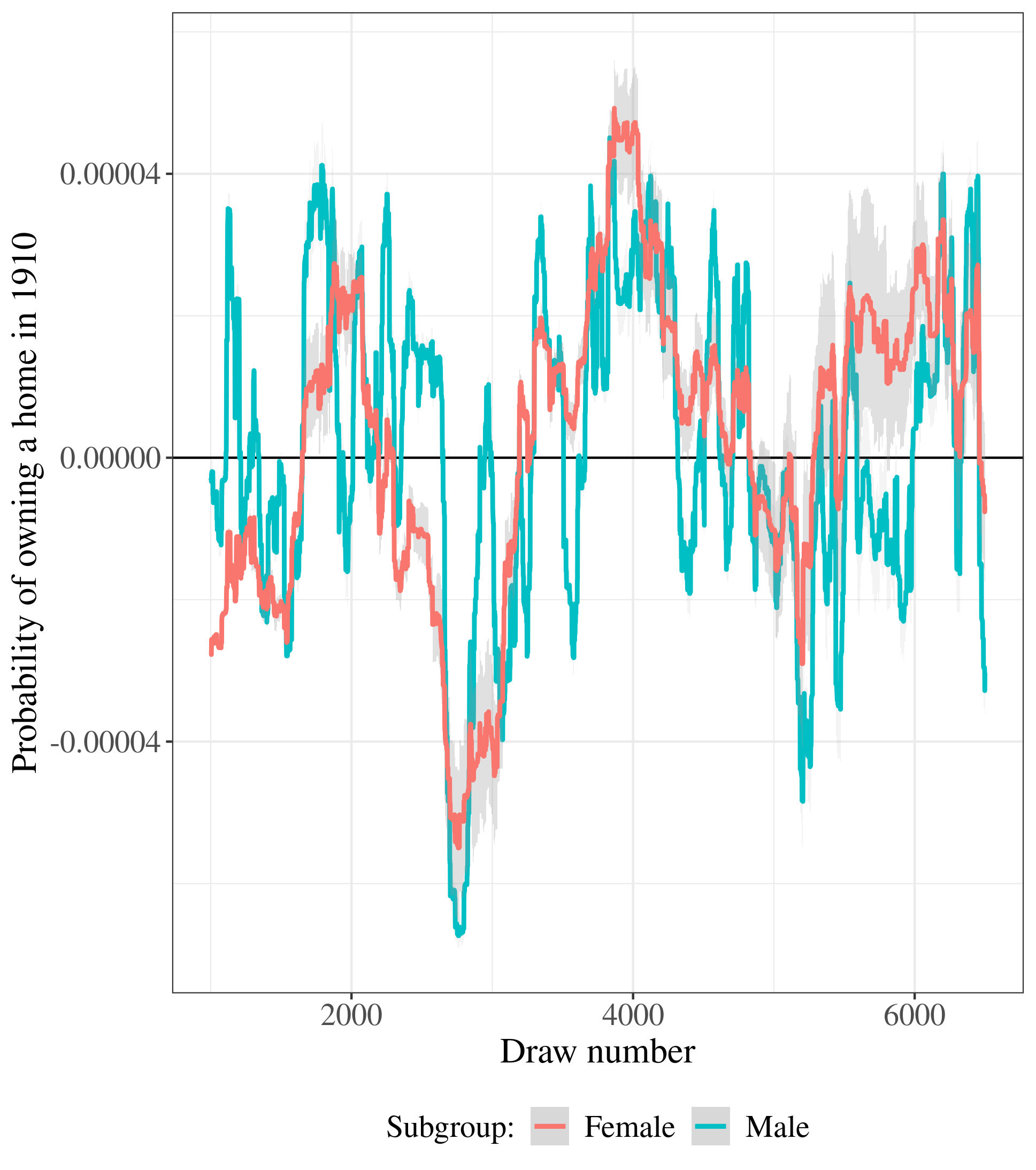}
		\caption{Marginal treatment effect function}
	\end{subfigure}
	\caption{Estimated average dose-response (a) and marginal treatment effect (b) on the probability of owning a home. See notes to Figure \ref{sale-plots}.\label{home-plots}} 
\end{figure}

\newpage
\begin{singlespace}
	\bibliographystyle{chicago}
	\bibliography{references}
\end{singlespace}

\itemize